\documentclass[journal,12pt,draftcls,peerreview,onecolumn,twoside]{IEEEtran}
\usepackage{latexsym}
\usepackage[psamsfonts]{amssymb}
\usepackage{cite}
\ifCLASSINFOpdf
\else
\usepackage[dvips]{graphicx}
\fi
\usepackage[cmex10]{amsmath}
\title{A Comprehensive Survey of Potential Game Approaches to Wireless Networks}
\author{
\IEEEauthorblockN{
\normalsize Koji Yamamoto%\IEEEauthorrefmark{1}
}
\IEEEauthorblockA{%\IEEEauthorrefmark{1}
\\ \small Graduate School of Informatics, Kyoto University\\
Yoshida-honmachi, Sakyo-ku, Kyoto 606-8501, Japan\\
E-mail: kyamamot@i.kyoto-u.ac.jp
}
}

\def\indicator#1{\mathop{\mathbbm{1}}\nolimits_{{#1}}}
\def\sgn{\mathop{\mathrm{sgn}}\nolimits}
\def\dd{{\rm d}}

\usepackage{mathtools}
\usepackage{tabularx}
\usepackage{bbm}
\usepackage{subfig}
\usepackage{booktabs}

\newcommand{\bm}[1]{\boldsymbol #1}
\newtheorem{dfn}{Definition}
\newtheorem{theorem}{Theorem}
\newtheorem{example}{Example}
\def\br{\mathop{\mathrm{BR}}\nolimits}
\def\cost{\mathop{\mathrm{cost}}\nolimits}

\begin{document}

\maketitle

\newcounter{gamenombre}
\renewcommand{\thegamenombre}{\arabic{gamenombre}}
\newcommand{\game}[1][]{\refstepcounter{gamenombre}#1 \thegamenombre}
\newcommand{\gameref}[1]{{\ref{#1}}}

\begin{abstract}
Potential games form a class of non-cooperative games where unilateral improvement dynamics are guaranteed to converge in many practical cases. The potential game approach has been applied to a wide range of wireless network problems, particularly to a variety of channel assignment problems. In this paper, the properties of potential games are introduced, and games in wireless networks that have been proven to be potential games are comprehensively discussed.
\end{abstract}
%\end{summary}

\section{Introduction}

The broadcast nature of wireless transmissions causes co-channel interference and channel contention, which can be viewed as interactions among transceivers. Interactions among multiple decision makers can be formulated and analyzed using a branch of applied mathematics called game theory \cite{Myerson1991,Fudenberg1991}. Game-theoretic approaches have been applied to a wide range of wireless communication technologies, including transmission power control for code division multiple access (CDMA) cellular systems \cite{Saraydar2002} and cognitive radios \cite{Neel2002game}. For a summary of game-theoretic approaches to wireless networks, we refer the interested reader to \cite{MacKenzie2006,Lasaulce2011,Han2011,Tembine2012,Lasaulce2009}. Application-specific surveys of cognitive radios and sensor networks can be found in \cite{Wang2010,Liu2010,Xu2013CSTO,Shi2012,Gavrilovska2013,Srivastava2005}.

In this paper, we focus on potential games \cite{Monderer1996}, which form a class of strategic form games with the following desirable properties:
\begin{itemize}
 \item The existence of a Nash equilibrium in potential games is guaranteed in many practical situations \cite{Monderer1996} (Theorems \ref{th:existence_finite} and \ref{th:existence_infinite} in this paper), but is not guaranteed for general strategic form games. Other classes of games possessing Nash equilibria are summarized in \cite[\S2.2]{Lasaulce2011} and \cite[\S3.4]{Han2011}.
 \item Unilateral improvement dynamics in potential games with finite strategy sets are guaranteed to converge to the Nash equilibrium in a finite number of steps, i.e., they do not cycle \cite{Monderer1996} (Theorem \ref{eq:converge_in_finite_steps} in this paper). As a result, learning algorithms can be systematically designed.
\end{itemize}
A game that does not have these properties is discussed in Example~\ref{example2} in Section~\ref{sec:game}.

We provide an overview of problems in wireless networks that can be formulated in terms of potential games. We also clarify the relations among games, and provide simpler proofs of some known results. Problem-specific learning algorithms \cite{Lasaulce2011,Tembine2012} are beyond the scope of this paper.

\begin{table*}[!t]
 \centering
 \caption{Games discussed in this paper.}
 \begin{tabular}{llll}
  \toprule
  Section & System model & Strategy & Payoff \\
  \midrule
  \ref{ssec:nie} & Fig.~\ref{fig:system_models}\subref{fig:model2} & Channel & Interference power \\
  \ref{sec:iSINR} & Fig.~\ref{fig:system_models}\subref{fig:model2} & Channel & SINR or Shannon capacity \\
  \ref{ssec:number_interference} & Fig.~\ref{fig:system_models}\subref{fig:model2} & Channel & Number of interference signals \\
  \ref{sec:neel} & Figs.~\ref{fig:system_models}\subref{fig:model3} and \ref{fig:system_models}\subref{fig:model4} & Channel & Interference power \\
  \ref{eq:neel_capacity} & Figs.~\ref{fig:system_models}\subref{fig:model3} and \ref{fig:system_models}\subref{fig:model4} & Channel & SINR or Shannon capacity \\
  \ref{sec:min_number} & Fig.~\ref{fig:system_models}\subref{fig:model5} & Channel & Number of interference signals \\
  \ref{sec:ca_aloha} & Fig.~\ref{fig:system_models}\subref{fig:model5} & Channel & Successful access probability or throughput \\
  \ref{sec:alpha}    & Fig.~\ref{fig:system_models}\subref{fig:model5} & Transmission probability & Successful access probability or throughput \\
  \ref{sec:tpc}  & Fig.~\ref{fig:system_models}\subref{fig:model1} & Transmission power & Throughput or Shannon capacity \\
  \ref{sec:topology} & Fig.~\ref{fig:system_models}\subref{fig:model3} & Transmission power & Connectivity \\
  \ref{sec:flow}     & Fluid network & Amount of traffic  & Congestion cost \\
  \ref{sec:mm1}      & M/M/1 queue & Arrival rate       & Trade-off between throughput and delay \\
  \ref{sec:location} & Mobile sensors & Location & Connectivity or coverage\\
  \ref{sec:immobile_sensor} & Immobile sensors & Channel & Coverage \\
  \bottomrule
 \end{tabular}
 \label{tab:system_models}
\end{table*}

The remainder of this paper is organized as follows: In Sections~\ref{sec:game}, \ref{sec:pot}, and \ref{sec:learning}, we introduce strategic form games, potential games, and learning algorithms, respectively. We then discuss various potential games in Sections \ref{ssec:nie} to \ref{sec:immobile_sensor}, as shown in Table \ref{tab:system_models}. Finally, we provide a few concluding remarks in Section \ref{sec:conclusion}.

\newsavebox{\boxa}

\sbox{\boxa}{
\includegraphics[scale=.5]{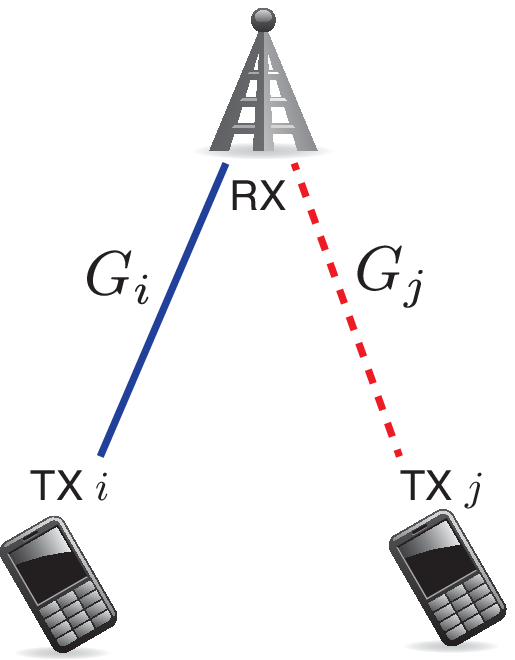}
}

\newsavebox{\boxc}

\sbox{\boxc}{
\includegraphics[scale=.5]{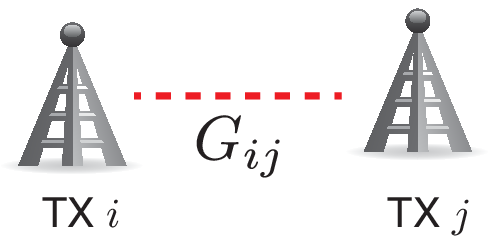}
}

\newsavebox{\boxd}

\sbox{\boxd}{
\includegraphics[scale=.5]{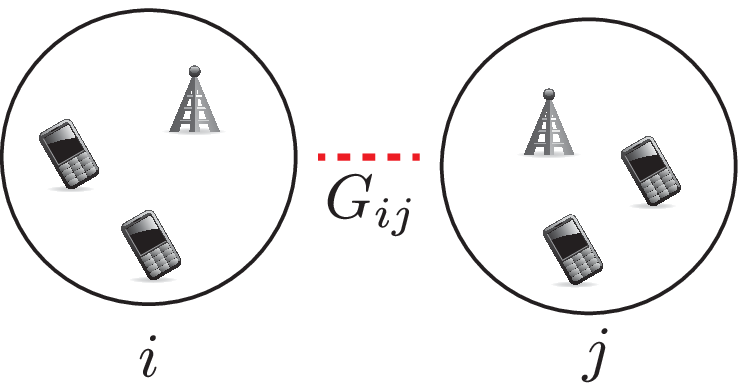}
}

\newsavebox{\boxe}

\sbox{\boxe}{
\includegraphics[scale=.5]{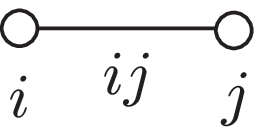}
}

\begin{figure*}[!t]
 \centering
 \subfloat[Multiple-access channel.]{\usebox{\boxa}\label{fig:model1}}
 \hfil
 \subfloat[TX-RX pairs \break ($G_{ij} \neq G_{ji}$).]{\includegraphics[scale=.5]{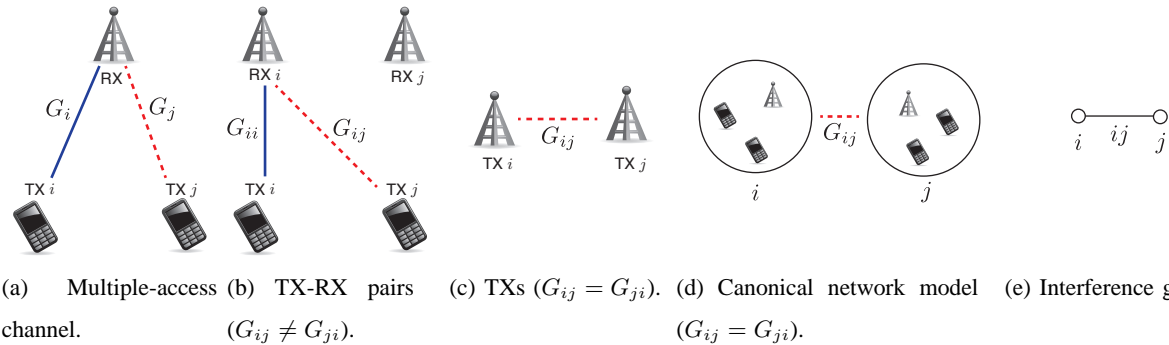}\label{fig:model2}}
 \hfil
 \subfloat[TXs ($G_{ij} = G_{ji}$).]{
 \vbox to \ht\boxa{\hsize\wd\boxc\vfil%
 \usebox{\boxc}\label{fig:model3}%
 \vfil}}
 \hfil
 \subfloat[Canonical network model \allowbreak ($G_{ij} = G_{ji}$).]{
 \vbox to \ht\boxa{\hsize\wd\boxd\vfil%
 \usebox{\boxd}\label{fig:model4}%
 \vfil}}
 \hfil
 \subfloat[Interference graph.]{
 \vbox to \ht\boxa{\hsize\wd\boxc\vfil%
 \usebox{\boxe}\label{fig:model5}%
 \vfil}}
 \caption{System models. Straight blue lines represent communication channels of player $i$ and red dashed lines represent interference channels to player $i$.}
 \label{fig:system_models}
\end{figure*}

The notation used here is shown in Table \ref{210446_20Aug14}. Unless the context indicates otherwise, sets of strategies are denoted by calligraphic uppercase letters, e.g., $ \mathcal{ A }_i $, strategies are denoted by lowercase letters, e.g., $ a_i \in \mathcal{ A }_i $, and tuples of strategies are denoted by boldface lowercase letters, e.g., $ \bm{ a } $. Note that $ a_i $ is a scalar variable when $ \mathcal{ A }_i $ is a set of scalars or indices, $ a_i $ is a vector variable when $ \mathcal{ A }_i $ is a set of vectors, and $ a_i $ is a set variable when $ \mathcal{ A }_i $ is a collection of sets.

We use $\mathbb{R}$ to denote the set of real numbers, $\mathbb{R}_{+}$ to denote the set of nonnegative real numbers, $\mathbb{R}_{++}$ to denote the set of positive real numbers, and $ \mathbb{ C } $ to denote the set of complex numbers. The cardinality of set $\mathcal{A}$ is denoted by $ \lvert \mathcal{A} \rvert $. The power set of $\mathcal{A}$ is denoted by $2^\mathcal{A} $. Finally, $\indicator{\mathit{condition}}$ is the indicator function, which is one when $\mathit{condition}$ is true and is zero otherwise.

We treat many system models, as shown in Fig.~\ref{fig:system_models}. In multiple-access channels, as shown in Fig.~\ref{fig:system_models}\subref{fig:model1}, multiple transmitters (TXs/users/mobile stations/terminals) transmit signals to a single receiver (RX/base station (BS)/access point (AP)). In Fig.~\ref{fig:system_models}\subref{fig:model1}, $G_{i}$ represents the link gain from TX $i$ to the RX.

In a network model consisting of TX-RX pairs, as shown in Fig.~\ref{fig:system_models}\subref{fig:model2}, each TX $ i $ transmits signals to RX $ i $. In this case, $ G_{ ij } \neq G_{ ji } $. In a network model consisting of TXs shown in Fig.~\ref{fig:system_models}\subref{fig:model3}, each TX (BS/AP/transceiver/station/terminal/node) interferes with others. In this model, $ G_{ ij } = G_{ ji } $. A ``canonical network model'' \cite{Babadi2010}, shown in Fig.~\ref{fig:system_models}\subref{fig:model4}, consists of clusters that are spatially separated in order for $ G_{ij} = G_{ji} $ to hold. Note that these network models have been discussed in terms of graph structure in \cite{Peltomaki2012}.

We use $ ji $ to denote a directed link from TX $ i $ to TX $ j $ or cluster $ i $ to cluster $ j $. Let interference graph $ ( \mathcal{ I }, \mathcal{ E } ) $ be an undirected graph, where the set of vertices $ \mathcal{ I } = \{ 1, 2, \dots \}$ corresponds to TXs or clusters, and $ i $ interferes with $ j $ if $ ji \in \mathcal{ E }$, as shown in Fig.~\ref{fig:system_models}\subref{fig:model5}, i.e., $ \mathcal{ E } \coloneqq \{\, ji \mid G_{ji} P > T \,\} $ where $ P $ is the transmission power level for every TX and $ T $ is a threshold of the received power. Note that in undirected graph $ ( \mathcal{ I }, \mathcal{ E } ) $, $ ji \in \mathcal{ E } \Leftrightarrow ij \in \mathcal{ E } $, for every $ \{ i, j \} \subset \mathcal{ I } $. We denote the neighborhood of $i$ in graph $ ( \mathcal{ I }, \mathcal{ E } ) $ by $ \mathcal{ I }_i \coloneqq \{\, j \in \mathcal{ I } \setminus \{ i \} \mid ji \in \mathcal{ E } \,\}$. We also define $ \mathcal{ I }_i^{ c_i }(\bm{ c }) \coloneqq \{\, j \in \mathcal{ I }_i \mid c_j = c_i \,\}$, then, $ \lvert \mathcal{ I }_i^{ c_i }(\bm{ c }) \rvert = \sum_{ j \in \mathcal{ I }_i } \indicator{ c_j = c_i } = \sum_{ j \neq i } \indicator{ c_j = c_i } \indicator{ ij \in \mathcal{ E } } $.

\begin{table}[!t]
 \centering
 \caption{Notation.}
\begin{tabularx}{\linewidth}{c|X}
 \toprule
 $ \mathcal{ G } $ & Strategic form game \\
 $ \mathcal{ I } $ & Finite set of players, $ \mathcal{ I } = \{ 1, 2, \dots, \lvert \mathcal{ I } \rvert \}$ \\
 $ \mathcal{ I }^f(\bm{a})$ & $ \coloneqq \{\, i \in \mathcal{ I } \mid f \in a_i \,\}$\\
 $\mathcal{ A }_i $ & Set of strategies for player $ i \in \mathcal{ I } $ \\
 $\mathcal{ A }$ & Strategy space, $ \prod_{ i \in \mathcal{ I } } \mathcal{ A }_i $ \\
 $ u_i $ & Payoff function for player $ i \in \mathcal{ I } $ \\
 $ \phi $   & Potential function \\
 $ \br_i $ & Best-response correspondence of player $i$ \\
 $ a_i $ & Strategy of player $ i $, $ a_i \in \mathcal{ A }_i $ \\
 $ \Delta ( \mathcal{ A }_i ) $ & Set of probability distributions over $ \mathcal{ A }_i $\\
 $ x_i $ & Mixed strategy, $ x_i \in \Delta ( \mathcal{ A }_i ) $\\
 $ \bm{x} $ & Mixed strategy profile, $ \bm{x} \in \prod_i \Delta ( \mathcal{ A }_i ) $\\
 \midrule
 $ G_i $ & Link gain between TX $i$ and a single isolated RX in Fig. \ref{fig:system_models}\subref{fig:model1} \\
 $ G_{ij} $ & Link gain between TX $j$ and RX $i$; $ G_{ ji } \neq G_{ ij }$ in Fig. \ref{fig:system_models}\subref{fig:model2}, and $ G_{ ji } = G_{ ij } $ in Figs. \ref{fig:system_models}\subref{fig:model3} and \ref{fig:system_models}\subref{fig:model4}\\
 \midrule
 $ ji $ & Directed link from $ i $ to $ j $ \\
 $ \mathcal{ E } $ & Set of edges in undirected graph \\
 $ \mathcal{ I }_{ i } $ & $ \coloneqq \{\, j \in \mathcal{ I } \setminus \{ i \} \mid ji \in \mathcal{ E }\,\}$. Neighborhood in graph $ ( \mathcal{ I }, \mathcal{ E } ) $ \\
 $ \mathcal{ I }_{ i }^{ c_i }(\bm{ c } ) $ & $ \coloneqq \{\, j \in \mathcal{ I }_i \mid c_j = c_i \,\}$.  \\
 \midrule
 $ N $ & Common noise power for every player \\
 $ N_i $ & Noise power at RX $i$ \\
 $ N_i(c_i) $ & Noise power at RX $i$ in channel $c_i$ \\
 $ I_i ( \bm{ c } ) $ & Interference power at RX $i$ at channel arrangement $ \bm{ c } $\\
 $ \mathcal{C}_i $ & Set of available channels for player $i$ \\
 $ c_i\ (\in \mathcal{C}_i) $ & Channel of player $i$ \\
 $ \bm{ c } $ & $ \coloneqq ( c_i )_{ i \in \mathcal{ I } } \in \prod_i \mathcal{ C }_i $ \\
 $ \mathcal{P}_i $ & Set of available transmission power levels for player $i$ \\
 $ p_i\ ( \in \mathcal{ P }_i ) $ & Transmission power level of player $i$ as a strategy \\
 $ \bm{ p } $ & $ \coloneqq ( p_i )_{ i \in \mathcal{ I } } \in \prod_i \mathcal{ P }_i $ \\
 $ P $ & Identical transmission power level for every player \\
 $ P_i $ & Transmission power level for player $ i $ as a constant \\
 $ \varGamma $ & Required signal-to-interference-plus-noise power ratio (SINR) \\
 \bottomrule
\end{tabularx}
 \label{210446_20Aug14}
\end{table}

\section{Game-theoretic Framework}
\label{sec:game}

We begin with the definition of a strategic form game and present an example of a game-theoretic formulation of a simple channel selection problem. Moreover, we discuss other useful concepts, such as the best response and Nash equilibrium. The analysis of Nash equilibria in the channel selection example reveals the potential presence of cycles in best-response adjustments.

\begin{dfn}
 A \textit{strategic (or normal) form game} is a triplet $ \mathcal{ G } \coloneqq (\mathcal{ I }, (\mathcal{A}_i)_{i \in \mathcal{ I }}, (u_i)_{i \in \mathcal{ I }})$, or simply $ \mathcal{ G } \coloneqq (\mathcal{ I }, (\mathcal{A}_i), (u_i))$, where $ \mathcal{ I } = \{ 1, 2, \dots, \lvert \mathcal{ I } \rvert \}$ is a finite set of \textit{players} (decision makers)\footnote{Infinite player (or non-atomic) potential games introduced in \cite{Sandholm2001,Sandholm2010} are beyond the scope of this paper. Infinite player potential games have been applied to BS selection games \cite{Shakkottai2007,Tran2010}.}, $\mathcal{A}_i$ is the set of \textit{strategies} (or actions) for player $i \in \mathcal{ I }$, and $u_i\colon \prod_{i \in \mathcal{ I }} \mathcal{A}_i \to \mathbb{R}$ is the \textit{payoff} (or utility) function of player $i \in \mathcal{ I }$ that must be maximized.

\end{dfn}

If $ \mathcal{S} \subseteq \mathcal{ I } $, we denote the Cartesian product $ \prod_{i \in \mathcal{S}} \mathcal{A}_i$ by $ \mathcal{A}_\mathcal{S}$. If $ \mathcal{S} = \mathcal{ I } $, we simply write $ \mathcal{A} $ to denote $ \mathcal{A}_\mathcal{ I } $, and $ \sum_{i} $ to denote $ \sum_{i \in \mathcal{ I }} $. When $ \mathcal{S} = \mathcal{ I }\setminus \{ i \} $, we let $ \mathcal{A}_{-i} $ denote $ \mathcal{A}_{\mathcal{ I } \setminus \{ i \} }, $ and $ \sum_{j \neq i} $  denote $ \sum_{ j \in \mathcal{ I } \setminus \{ i \} } $. For $ a_i \in \mathcal{A}_i $, $ \bm{ a }_\mathcal{ S } = ( a_i )_{ i \in \mathcal{ S } } \in \mathcal{ A }_\mathcal{ S } $, $ \bm{a} = (a_i, \bm{a}_{-i}) = (a_1,\ldots,a_{ \lvert \mathcal{ I } \rvert }) \in \mathcal{A} $, and $ \bm{a}_{-i} = ( a_1, \ldots, a_{i - 1}, a_{i + 1}, \ldots, a_{ \lvert \mathcal{ I } \rvert } ) \in \mathcal{A}_{-i} $.

\begin{example}
\label{example1}

Consider a channel selection problem in the TX-RX pair model shown in Fig.~\ref{fig:system_models}\subref{fig:model2}. Each TX-RX pair is assumed to select its channel in a decentralized manner in order to minimize the received interference power.

The channel selection problem can be formulated as a strategic form game $ \mathcal{G}\game[\label{g:u1}] \coloneqq ( \mathcal{ I }, ( \mathcal{C}_i) , ( u\gameref{g:u1}_i ) ) $. The elements of the game are as follows: the set of players $ \mathcal{ I } $ is the set of TX-RX pairs. The strategy set for each pair $ i $, $ \mathcal{ C }_i $ is the set of available channels. The received interference power at RX $ i \in \mathcal{ I } $ is determined by a combination of channels $ \bm{ c } =  ( c_i )_{ i \in \mathcal{ I } } \in \mathcal{ C } = \prod_i \mathcal{ C }_i $, where
\begin{align}
 I_i ( \bm{ c } ) &\coloneqq \sum_{ j \neq i } G_{ ij } P \indicator{ c_j = c_i }.
 \label{eq:interference}
 \end{align}
Let $ -I_i ( \bm{ c } )$ be the payoff function to be maximized, i.e.,
\begin{align}
 u\gameref{g:u1}_i ( \bm{ c } ) \coloneqq -I_i ( \bm{c} ) = - \sum_{ j \neq i } G_{ ij } P \indicator{ c_j = c_i }.
\end{align}
Note that $\mathcal{G}\gameref{g:u1}$ was introduced in \cite{Nie2006}, and we further discuss it in Example \ref{example2}.
\end{example}

\begin{dfn}
 The \textit{best-response correspondence}\footnote{A correspondence is a set-valued function for which all image sets are non-empty, e.g, \cite{Myerson1991,Lasaulce2011}.} (or simply, best response) $ \br_i \colon \mathcal{A}_{-i} \to 2^{\mathcal{A}_i}$ of player $i$ to strategy profile $\bm{a}_{-i}$ is the correspondence
 \begin{multline}
  \br_i(\bm{a}_{-i}) \\
  \coloneqq  \{\, a_i \in \mathcal{A}_i \mid u_i(a_i,\bm{a}_{-i}) \geq u_i (a_i',\bm{a}_{-i}), \forall a_i' \in \mathcal{A}_i \,\},
 \end{multline}
 or equivalently, $ \br_i(\bm{a}_{-i}) \coloneqq \arg\max_{a_i \in \mathcal{A}_i} u_i( a_i,\bm{a}_{-i} ) $.

\end{dfn}

A fundamental solution concept for strategic form games is the Nash equilibrium:
\begin{dfn}
 A strategy profile $ \bm{a}^* = ( a_i^*, \bm{a}_{-i}^* ) \in \mathcal{A} $ is a pure-strategy \textit{Nash equilibrium} (or simply a Nash equilibrium) of game $ ( \mathcal{ I }, (\mathcal{A}_i), (u_i) ) $ if
 \begin{align}
  u_i(a_i^*,\bm{a}_{-i}^*) \geq u_i(a_i,\bm{a}_{-i}^*), \label{eq:ne}
 \end{align}
 for every $ i \in \mathcal{ I } $ and $ a_i \in \mathcal{A}_i $; equivalently,
  $a_i^* \in \br_i(\bm{a}_{-i}^*)$ for every $ i \in \mathcal{ I } $.
That is, $a_i^*$ is a solution to the optimization problem $ \max_{ a_i \in \mathcal{A}_i } u_i (a_i, \bm{a}_{-i}^*)$.
\end{dfn}

At the Nash equilibrium, no player can improve his/her payoff by adopting a different strategy \textit{unilaterally}; thus, no player has an incentive to unilaterally deviate from the equilibrium. The Nash equilibrium is a proper solution concept; however, the existence of a pure-strategy Nash equilibrium is not necessarily guaranteed, as shown in the next example.

\begin{figure}[t]
 \centering
 \includegraphics[width=\linewidth]{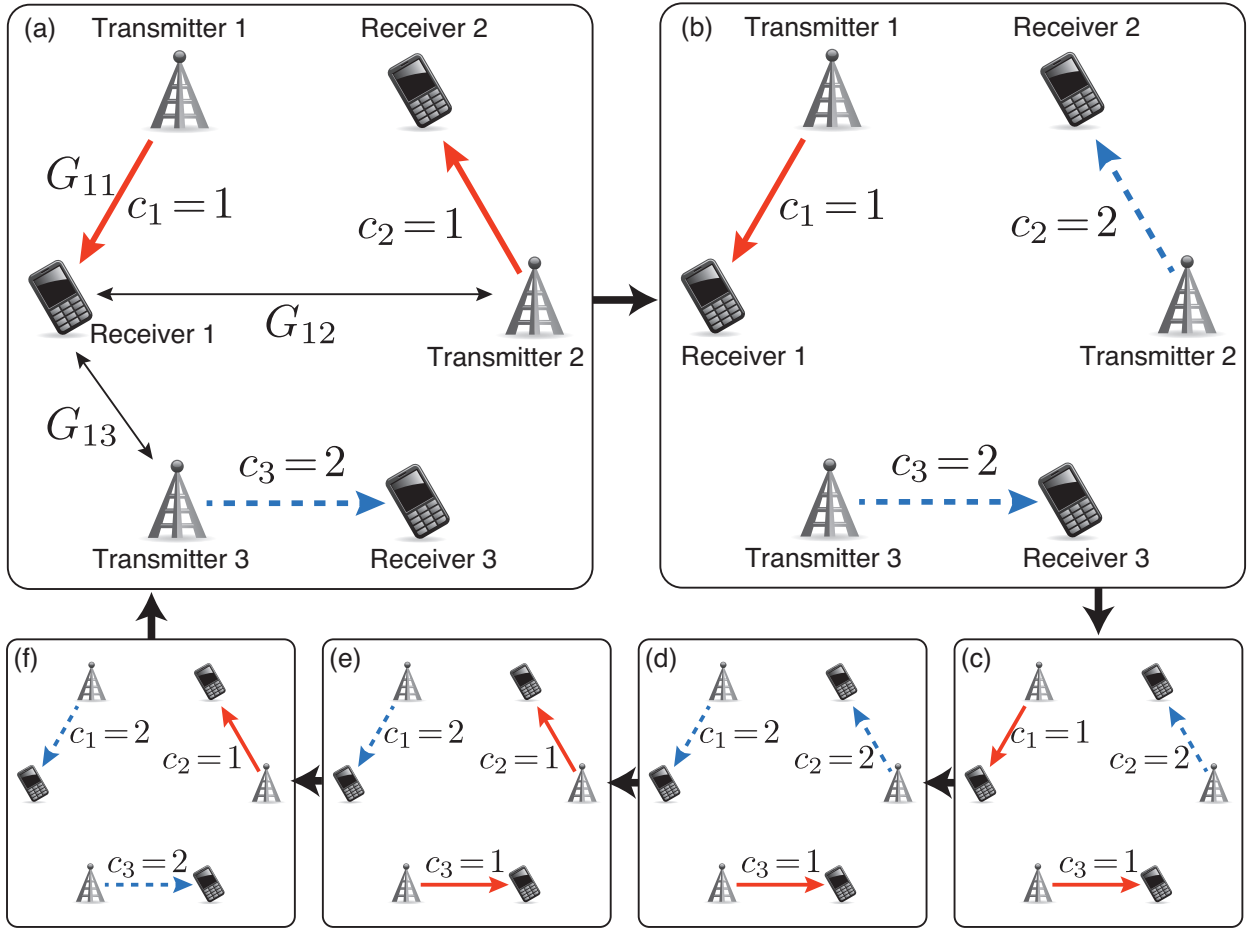}
 \caption{Arrangement used in Example \ref{example2}. A cycle results from the best-response adjustment.}
 \label{fig:loop}
\end{figure}

\begin{example}
\label{example2}
Consider $\mathcal{G}\gameref{g:u1}$ and the arrangement shown in Fig.~\ref{fig:loop}, i.e., $ \mathcal{ I } = \{ 1, 2, 3 \}$, $ \mathcal{ C }_i = \{ 1, 2 \}$ for every $i$, and $G_{13} > G_{12}$, $G_{21}>G_{23}$, and $G_{32} > G_{31}$\footnote{This setting is essentially the same as that used in \cite{Gallego2012},\,\allowbreak\cite{Menon2006},\,\allowbreak\cite[Example 4.17]{Neel2006-216analysis},\,\allowbreak\cite{Neel2007-5synthetic}.}. The game does not have a Nash equilibrium, i.e., for every channel allocation, at least one pair has an incentive to change his/her channel. The details are as follows: when all players choose the same channel, e.g., $ (c_1, c_2, c_3) = (1, 1, 1) $, every player has an incentive to change his/her channel because $ \br_i(\bm{c}_{-i}) = \{ 2 \} $ for all $i$; thus, it is not in Nash equilibrium. On the contrary, when two players choose the same channel, and the third player chooses a different channel, e.g., $ ( c_1, c_2, c_3 ) = ( 1, 1, 2 ) $, as shown in Fig.\ \ref{fig:loop}(a), $ \br_{2}(\bm{c}_{-2}) = \{ 2 \} $, i.e., pair 2 has an incentive to change its channel $a_2$ from 1 to 2, and (\ref{eq:ne}) does not hold. Because of the symmetry property of the arrangement in Fig.~\ref{fig:loop}, every strategy profile does not satisfy (\ref{eq:ne}). Furthermore, the best-response channel adjustments, which will be formally discussed in Section \ref{sec:learning}, cycle as $(1, 1, 2)$, $(1, 2, 2)$, $(1, 2, 1)$, $(2, 2, 1)$, $(2, 1, 1)$, $(2, 1, 2)$, and $(1, 1, 2)$, as shown in Figs.~\ref{fig:loop}(a-f).
\end{example}
The channel allocation game $ \mathcal{G}\gameref{g:u1} $ is discussed further in Section \ref{ssec:nie}.

\section{Potential Games}
\label{sec:pot}

We state key definitions and properties of potential games in Section \ref{ssec:properties}, show how to identify and design exact potential games in Sections \ref{ssec:identification_exact} and \ref{ssec:design}, and show how to identify ordinal potential games in Section \ref{ssec:identification_ordinal}.

\subsection{Definitions and Properties of Potential Games}
\label{ssec:properties}

Monderer and Shapley \cite{Monderer1996} introduced the following classes of potential games\footnote{There are a variety of generalized concepts of potential games, e.g., generalized ordinal potential games \cite{Monderer1996}, best-response potential games \cite{Voorneveld2000}, pseudo-potential games \cite{Dubey2006}, near-potential games \cite{Candogan2013,Candogan2013a}, and state-based potential games \cite{Marden2012}. Applications of these games are beyond the scope of this paper.}:
\begin{dfn}
 A strategic form game $ (\mathcal{ I },(\mathcal{A}_i),(u_i))$ is an \textit{exact potential game} (EPG) if there exists an \textit{exact potential function} $\phi \colon \mathcal{A} \to \mathbb{R}$ such that
 \begin{align}
  u_i ( a_{i}, \bm{a}_{-i} ) - u_i ( a_{i}', \bm{a}_{-i} )
  &=
  \phi ( a_{i}, \bm{a}_{-i} ) - \phi ( a_{i}', \bm{a}_{-i} ), \label{eq:epg}
  \end{align}
 for every $i \in \mathcal{ I }$, $a_i, a_i' \in \mathcal{A}_i$, and $\bm{a}_{-i} \in \mathcal{A}_{-i}$.
\end{dfn}

\begin{dfn}
 A strategic form game $(\mathcal{ I },(\mathcal{A}_i),(u_i))$ is a \textit{weighted potential game} (WPG) if there exist a \textit{weighted potential function} $\phi \colon \mathcal{A} \to \mathbb{R}$ and a set of positive numbers $\{ \alpha_i \}_{i\in\mathcal{ I }}$ such that
\begin{align}
  u_i ( a_{i}, \bm{a}_{-i} ) - u_i ( a_{i}', \bm{a}_{-i} )
  =
  \alpha_i (\phi ( a_{i}, \bm{a}_{-i} ) - \phi ( a_{i}', \bm{a}_{-i} ) ), \label{eq:wpg}
\end{align}
 for every $ i \in \mathcal{ I }$, $a_i, a_i' \in \mathcal{A}_i$, and $\bm{a}_{-i} \in \mathcal{A}_{-i}$.
\end{dfn}

\begin{dfn}
 A strategic form game $(\mathcal{ I },(\mathcal{A}_i),(u_i))$ is an \textit{ordinal potential game} (OPG) if there exists an \textit{ordinal potential function} $\phi \colon \mathcal{A} \to \mathbb{R}$ such that
\begin{multline}
 \sgn(u_i(a_i, \bm{a}_{-i}) - u_i(a_i', \bm{a}_{-i})) \\
 =\sgn(\phi(a_i, \bm{a}_{-i}) - \phi(a_i', \bm{a}_{-i})),
\end{multline}
 for every $ i \in \mathcal{ I }$, $a_i, a_i' \in \mathcal{A}_i$, and $\bm{a}_{-i} \in \mathcal{A}_{-i}$, where $\sgn(\cdot)$ denotes the sign function.
\end{dfn}

Although the potential function $ \phi $ is independent of the indices of the players, $ \phi $ reflects any unilateral change in any payoff function $u_i$ for every player $i$.

Since an EPG is a WPG and a WPG is an OPG \cite{Monderer1996,Voorneveld2000}, the following properties of OPGs are satisfied by EPGs and WPGs.

\begin{theorem}[Existence in finite OPGs]
 \label{th:existence_finite}
Every OPG with finite strategy sets possesses at least one Nash equilibrium \cite[Corollary 2.2]{Monderer1996}.
\end{theorem}

\begin{theorem}[Existence in infinite OPGs]\label{th:existence_infinite}
In the case of infinite strategy sets, every OPG with compact strategy sets and continuous payoff functions possesses at least one Nash equilibrium \cite[Lemma 4.3]{Monderer1996}.
\end{theorem}

\begin{theorem}[Uniqueness]\label{th:unique}
 Every OPG with a compact and convex strategy space, and a strictly concave and continuously differentiable potential function possesses a unique Nash equilibrium \cite[Theorem 2]{Neyman1997},\,\allowbreak\cite{Scutari2006}.
\end{theorem}

The most important property of potential games is \textit{acyclicity}, which is also referred to as the finite improvement property.

\begin{dfn}[Finite improvement property \cite{Monderer1996}]
A \textit{path} in $(\mathcal{ I }, (\mathcal{A}_i), (u_i) )$ is a sequence $(\bm{a}[0], \bm{a}[1], \ldots)$ such that for every integer $ k \geq 1 $, there exists a unique player $i$ such that $a_i[k] \neq a_i[k\!-\!1] \in \mathcal{A}_i$ while $\bm{a}_{-i}[k] = \bm{a}_{-i}[k\!-\!1]$.
 $(\bm{a}[0], \bm{a}[1], \ldots)$ is an \textit{improvement path} if, for every $k \geq 1$, $u_i(\bm{a}[k]) > u_i(\bm{a}[k\!-\!1])$, where $i$ is the unique deviator at step $k$. $(\mathcal{ I }, (\mathcal{A}_i), (u_i) )$ has the \textit{finite improvement property (FIP)} if every improvement path is finite.
\end{dfn}

\begin{theorem}
 \label{eq:converge_in_finite_steps}
 Every OPG with finite strategy sets has the FIP \cite[Lemma 2.3]{Monderer1996}; that is, unilateral improvement dynamics are guaranteed to converge to a Nash equilibrium in a finite number of steps.
\end{theorem}

\subsection{Identification of Exact Potential Games}
\label{ssec:identification_exact}

The definition of an EPG utilizes a potential function (\ref{eq:epg}). Sometimes, however, it is beneficial to know if a given game is an EPG independently of its potential function. The following properties of EPGs and classes of games known to be EPGs are useful for the identification and derivation of potential functions. Note that each EPG has a unique exact potential function except for an additive constant \cite[Lemma 2.7]{Monderer1996}.

\begin{theorem}
Let $ ( \mathcal{ I }, ( \mathcal{A}_i ), ( u_i ) ) $ be a strategic form game where strategy sets $\mathcal{A}_i $ are intervals of real numbers and payoff functions $ u_i $ are twice continuously differentiable. Then, the game is an EPG if and only if
\begin{align}
 \frac{\partial^2 u_i ( \bm{ a } ) }{\partial a_i\, \partial a_j} &=
 \frac{\partial^2 u_j ( \bm{ a } ) }{\partial a_i\, \partial a_j},
 \label{eq:twice_difference}
\end{align}
for every $ i, j \in \mathcal{ I } $ \cite[Theorem 4.5]{Monderer1996}.
\end{theorem}

\begin{theorem}
 \label{th:linear}
 Let $(\mathcal{ I },(\mathcal{A}_i),(u_{i,1}))$ and $(\mathcal{ I },(\mathcal{A}_i),(u_{i, 2}))$ be EPGs with potential functions $\phi_1(\bm{a})$ and $\phi_2(\bm{a})$, respectively. Furthermore, let $\alpha,\ \beta \in \mathbb{R}$.  Then, $(\mathcal{ I },(\mathcal{A}_i),(\alpha u_{i, 1} + \beta u_{i, 2}))$ is an EPG with potential function $\alpha \phi_1(\bm{a}) + \beta \phi_2(\bm{a})$  \cite{Facchini1997}.
\end{theorem}

\subsubsection{Coordination-dummy Games}
\label{sssec:coord-dummy}

 If $ u_i( \bm{ a } ) = u( \bm{ a } )$ for all $ i \in \mathcal{ I } $, where $ u \colon \mathcal{ A } \to \mathbb{ R } $, the game $ ( \mathcal{ I }, ( \mathcal{ A }_i ), ( u ) ) $ is called a \textit{coordination game}\footnote{The term ``coordination game'' is also used to describe games where players receive benefits when they choose the same strategy \cite{Cooper1999}.} or an \textit{identical interest game}, and $ u $ is called a coordination function \cite{Facchini1997}.

 If $ u_i( \bm{ a } ) = d_i( \bm{ a }_{ -i } )$ for all $ i \in \mathcal{ I } $, where $ d_i \colon \mathcal{ A }_{ -i } \to \mathbb{ R } $, the game  $ ( \mathcal{ I }, ( \mathcal{ A }_i ), ( d_i ) ) $ is called a \textit{dummy game}, and $ d_i $ is called a dummy function \cite{Facchini1997}.

 If $ u_i( \bm{ a } ) = s_i( a_i )$ for all $ i \in \mathcal{ I } $, where $ s_i \colon \mathcal{ A }_{ i } \to \mathbb{ R } $, the game  $ ( \mathcal{ I }, ( \mathcal{ A }_i ), ( s_i ) ) $ is called a \textit{self-motivated game}, and $ s_i $ is called a self-motivated function \cite{Neel2004}.

\begin{theorem}
 \label{th:coord-dummy}
 $(\mathcal{ I },(\mathcal{A}_i),(u_i))$ is an EPG if and only if there exist functions $u\colon \mathcal{A} \to \mathbb{R}$ and $d_i \colon \mathcal{A}_{-i} \to \mathbb{R}$ such that
 \begin{align}
  u_i(a_i, \bm{a}_{-i}) = u(a_i, \bm{a}_{-i}) + d_i(\bm{a}_{-i}),
 \end{align}
 for every $ i \in \mathcal{ I } $ \cite{Slade1994,Facchini1997}. This game is said to be a \textit{coordination-dummy game}.  The potential function of this game is $ \phi(\bm{a}) = u(\bm{a})$.
\end{theorem}

\begin{example}
 From Theorem \ref{th:coord-dummy}, any identical interest game is an EPG. Almost all games found in studies applying identical interest games \cite{Song2008joint,He2008,Lu2011,Duarte2012,Canales2012,Ortin2013,Bennis2013} have the form of game $ \mathcal{G}\game[\label{g:coord}] \coloneqq ( \mathcal{ I }, ( \mathcal{ A }_i ), ( u\gameref{g:coord}_i ) ) $, where
\begin{align}
 u\gameref{g:coord}_i( \bm{ a } ) &\coloneqq \sum_{ j } f_j ( \bm{ a } ),
\end{align}
 for every $ i \in \mathcal{ I } $ and  $ f_j ( \bm{ a } ) $ is a performance indicator of player $ j $, e.g., $f_j (\bm{a})$ is the individual throughput and $ u\gameref{g:coord}_i( \bm{ a } ) $ is the aggregated throughput of all players \cite{Song2008joint}. Note that in most of these works, $ \mathcal{G} \gameref{g:coord} $ is used for comparison with other games.

\end{example}

\begin{example}
Closely related to $ \mathcal{G} \gameref{g:coord} $, the form of game $\mathcal{G}\game[\label{eq:coord_graph}]$ with payoff
\begin{align}
 u\gameref{eq:coord_graph}_i ( \bm{ a } ) &\coloneqq f_i( a_i, \bm{a}_{\mathcal{ I }_i } ) + \sum_{j \in \mathcal{ I }_i } f_j ( a_j, \bm{ a }_{ \mathcal{I}_j } ),
\end{align}
where $ f_i \colon \mathcal{ A }_i \times \mathcal{ A }_{ \mathcal{ I }_i } \to \mathbb{ R } $, is found in many scenarios: data stream control in multiple-input and multiple-output (MIMO) \cite{Arslan2007equilibrium}, channel assignment \cite{Xu2012-42STSP}, joint power, channel and BS assignment \cite{Singh2013}, joint power and user scheduling \cite{Zheng2014WCOM}, BS selection \cite{Du2014}, and BS sleeping \cite{Zheng2015WCOM}. Note that $ \mathcal{G} \gameref{eq:coord_graph} $ is not an identical interest game, but can be seen as $ \mathcal{G} \gameref{g:coord} $ on graphs, where the performance indicator of player $ i $ is a function of strategies of its neighbors, i.e., $ f_i \colon \mathcal{ A }_i \times \mathcal{ A }_{ \mathcal{ I }_i } \to \mathbb{ R } $, and the sum of the performance indicators of player $ i $ and neighbors $ \mathcal{ I }_i $ is set for the payoff function of player $ i $. It can be easily proved that $ \mathcal{G} \gameref{eq:coord_graph} $ is an EPG with potential
\begin{align}
 \phi\gameref{eq:coord_graph} ( \bm{ a } ) &= \sum_{ i } f_i (a_i, \bm{ a }_{ \mathcal{ I }_i } ).
\end{align}

\end{example}

\subsubsection{Bilateral Symmetric Interaction Games}

 A strategic form game $\mathcal{G}\game[\label{eq:bsi}] \coloneqq (\mathcal{ I },(\mathcal{A}_i),(u\gameref{eq:bsi}_i ))$ is called a \textit{bilateral symmetric interaction (BSI) game} if there exist functions $w_{ij}\colon\mathcal{A}_i \times \mathcal{A}_j \to \mathbb{R}$ and $s_i\colon\mathcal{A}_i \to \mathbb{R}$ such that
\begin{align}
 u\gameref{eq:bsi}_i (\bm{a}) &= \sum_{j \neq i} w_{ij} (a_i,a_j) - s_i(a_i),
\end{align}
 where $w_{ij} (a_i, a_j) = w_{ji} (a_j, a_i)$ for every $(a_i, a_j) \in \mathcal{A}_i \times \mathcal{A}_j$ \cite{Ui2000}.

\begin{theorem}[\cite{Ui2000}]
 A BSI game $ \mathcal{G} \gameref{eq:bsi} $ is an EPG with potential function\footnote{$\sum_{i < j} = \sum_{\{ i, j \} \subseteq \mathcal{ I }} = \sum_{i = 1}^{ \lvert \mathcal{ I } \rvert } \sum_{ j = i + 1 }^{  \lvert \mathcal{ I } \rvert }$.}
 \begin{align}
 \phi\gameref{eq:bsi} (\bm{a})
  &= \frac{1}{2}\sum_{i} \sum_{j \neq i}w_{ij} (a_i, a_j) - \sum_{i} s_i(a_i) \nonumber \nonumber \\
 &= \sum_{i < j} w_{ij} (a_i, a_j) - \sum_{i} s_i(a_i).
\end{align}

\end{theorem}

\begin{example}
Consider a \textit{quasi-Cournot game} $ \mathcal{G}\game[\label{eq:cournot}] \coloneqq ( \mathcal{ I }, ( \mathcal{ A }_i ), ( u\gameref{eq:cournot}_i ) ) $ with a linear inverse demand function, where each player $ i \in \mathcal{ I } $ produces a homogeneous product and determines the output. Let $ \mathcal{A}_i = \mathbb{R}_{++} $ be a set of possible outputs. The payoff function of player $i$ is defined by
\begin{align}
 u\gameref{eq:cournot}_i (\bm{a}) &\coloneqq
 \left(
  \alpha - \beta \textstyle\sum_{j} a_j
 \right)
 a_i - \cost_i(a_i),
\end{align}
where $\alpha,\ \beta > 0$ and $\cost_i \colon \mathcal{A}_i \to \mathbb{R}$ is a differentiable cost function. Since
\begin{align}
 u\gameref{eq:cournot}_i (\bm{a}) &= \underbrace{\alpha a_i - \beta {a_i}^2  - \cost_i(a_i)}_{\text{self-motivated function}}
  - \underbrace{\beta \textstyle \sum_{ j \neq i } a_j \, a_i }_{ \text{BSI} },
\end{align}
$ \mathcal{G} \gameref{eq:cournot} $ is an EPG with potential
\begin{align}
 \phi\gameref{eq:cournot} (\bm{a}) &= \alpha \textstyle \sum_{i} a_i - \beta \sum_{i} {a_i}^2 - \sum_{i} c_i (a_i) - \beta \sum_{i<j} a_i\, a_j
\end{align}
\cite{Slade1994}. Further discussion can be found in \cite{Monderer1996,Ui2000}.

\end{example}

\subsubsection{Interaction Potential}

\begin{theorem}[\cite{Ui2000}]
 \label{th:interactionpotential}
A normal form game $ \mathcal{G}\game[\label{eq:interaction_payoff}] \coloneqq ( \mathcal{ I }, ( \mathcal{A}_i ) , ( u\gameref{eq:interaction_payoff}_i ) )$ is an EPG if and only if there exists a function $ \{\, \varPhi_\mathcal{ S } \mid \varPhi_\mathcal{ S } \colon \mathcal{ A }_\mathcal{ S } \to \mathbb{ R }, \mathcal{ S } \subseteq \mathcal{ I } \,\} $ (called an \textit{interaction potential}) such that
\begin{align}
 u\gameref{eq:interaction_payoff}_i ( \bm{a} ) &= \sum_{ \mathclap{ \mathcal{ S } \subseteq \mathcal{ I } : i \in \mathcal{ S } } } \varPhi_\mathcal{ S } ( \bm{ a }_\mathcal{ S } ),
 \label{eq:interaction_payoff_utility}
\end{align}
for every $ \bm{ a } \in \mathcal{A}$ and $i \in \mathcal{ I }$.
 The potential function is
\begin{align}
 \phi\gameref{eq:interaction_payoff} ( \bm{ a } ) &= \sum_{ \mathclap{ \mathcal{ S } \subseteq \mathcal{ I } } } \varPhi_\mathcal{ S } ( \bm{ a }_\mathcal{ S } ).
 \label{eq:interaction_potential}
\end{align}
\end{theorem}

\subsubsection{Congestion Games}
\label{sssec:congestion_game}

In congestion games (CGs), the payoff for using a resource (e.g., a channel or a facility) is a function of the number of players using the same resource. More precisely, CGs are defined as follows:

In the congestion model proposed by Rosenthal \cite{Rosenthal1973}, each player $ i $ uses a subset $ a_i $ of common resources $ \mathcal{ F }$, and receives resource-specific payoff $ w_f ( \lvert \mathcal{ I }^f ( \bm{ a } ) \rvert ) $ from resource $ f \in a_i $ according to the number of players using resource $ f $. Here, $ w_f \colon \{ 1, \dots, \lvert \mathcal{ I } \rvert \} \to \mathbb{ R } $,
$ \mathcal{ I }^f ( \bm{ a } ) \coloneqq \{\, i \in \mathcal{ I } \mid f \in a_i \,\} $ represents the set of players that use resource $ f $. Then, $ \lvert \mathcal{ I }^f ( \bm{ a } ) \rvert = \sum_{ i } \indicator{ f \in a_i } $.

A strategic form game $ \mathcal{G}\game[\label{eq:CG}] \coloneqq (\mathcal{ I }, ( \mathcal{ A }_i ), ( u\gameref{eq:CG}_i ) )$ associated with a congestion model, where $ \mathcal{ A }_i \subseteq 2^\mathcal{ F } $ and
\begin{align}
 u\gameref{eq:CG}_i ( \bm{ a } ) &\coloneqq \sum_{ \mathclap{ f \in a_i  } } w_f( \lvert \mathcal{ I }^f ( \bm{ a } ) \rvert ),
\end{align}
is called a CG.
Note that $ \mathcal{ A }_i $ is a collection of subsets of $ \mathcal{ F } $ and is not a set. Moreover, $ a_i \in \mathcal{ A }_i $ is a set, not a scalar quantity.
Note that a CG where the strategy of every player is a singleton, i.e.,
$ \mathcal{ A }_i \subseteq \mathcal{ F } $ and $ u\gameref{eq:CG}_i ( \bm{ a } ) = w_{a_i} ( \lvert \mathcal{ I }^{a_i} ( \bm{ a } ) \rvert ) $
is called a \textit{singleton CG}.

\begin{theorem}
 \label{th:congestion}
A CG $ \mathcal{G} \gameref{eq:CG} $ is an EPG with potential function
\begin{align}
 \phi\gameref{eq:CG} ( \bm{a} ) = \sum_{f \in \cup_{i} a_i } \left( \sum_{ k = 1 }^{ \lvert \mathcal{ I }^f ( \bm{ a } ) \rvert } w_f ( k )
 \right),
\end{align}
\cite[Theorem 3.1]{Monderer1996}\,\allowbreak\cite{Rosenthal1973}.
Furthermore, every EPG with finite strategy sets has an equivalent CG \cite[Theorem 3.2]{Monderer1996}.
\end{theorem}

Note that generalized CGs do not necessarily possess potential functions. For generalized CGs with potential, we refer the interested reader to \cite{Mavronicolas2007,Ackermann2009}. It was proved that CGs with player-specific payoff functions \cite{Milchtaich1996}, and those with resource-specific payoff functions and player-specific constants \cite{Mavronicolas2007}, have potential. CGs with linear payoff function on undirected/directed graphs has been discussed in \cite{Bilo2011Algorithmica}.

\subsection{Design of Payoff Functions}
\label{ssec:design}
In some scenarios, we can design payoff functions and assign them to players to ensure that the game is an EPG. Such approach is often applied in the context of cooperative control \cite{Marden2009}. These design methodologies can be used when we want to derive payoff functions from a given global objective so that the game with the designed payoff functions is an EPG with the global objective as the potential function. If the global objective is in the form of (\ref{eq:interaction_potential}), we can derive payoff functions by using (\ref{eq:interaction_payoff_utility}).

Otherwise, we can utilize many design rules: the equally shared rule, marginal contribution, and the Shapley values \cite{Shapley1952,Ui2000}. Since Marden and Wierman \cite{Marden2013} have already summarized these rules, we only present marginal contribution here.

Marginal contribution, or the \textit{wonderful life utility (WLU)} \cite{Wolpert1999}, is the following payoff function derived from the potential function:
\begin{align}
 u_i ( \bm{a} ) &= \phi ( \bm{a} ) - \phi ( \bm{a}_{-i} ),
 \label{eq:WLU1}
\end{align}
where $ \phi ( \bm{a}_{-i} ) $ is the value of the potential function in the absence of player $i$. The game with the WLU is an EPG with potential function $\phi$ \cite{Marden2013}.

When the potential function for each player is represented as the sum of functions $ f_i \colon \mathcal{ A } \to \mathbb{ R } $, i.e.,
$ \phi ( \bm{a} ) = \sum_{ j } f_j ( \bm{ a } )$ and
$ \phi ( \bm{a}_{-i} ) = \sum_{ j \neq i } f_j( \bm{ a }_{ -i } )$,
the WLU (\ref{eq:WLU1}) can be written as
\begin{align}
 u_i ( \bm{ a } ) &= \textstyle\sum_j f_j( \bm{ a } ) - \textstyle\sum_{ j \neq i } f_j( \bm{ a_{ -i } } ) \nonumber \\
 &= f_i ( \bm { a } ) - \textstyle\sum_{ j \neq i } (  f_j ( \bm{ a }_{ -i } ) - f_j( \bm{ a } ) ),
 \label{eq:WLU2}
\end{align}
where $ f_j ( \bm{ a }_{ -i } ) - f_j( \bm{ a } ) $ represents the loss to player $ j $ resulting from player $i$'s participation.

\begin{example}[Consensus game]
In the consensus problem \cite{Tsitsiklis1986}, each player $ i $ adjusts $ a_i $ and tries to reach $ a_1 = a_2 = \dots = a_{ \lvert \mathcal{ I } \rvert } $.

Marden et al.\@ \cite{Marden2009} considered the global objective
\begin{align}
 \phi\gameref{eq:marden_consensus} ( \bm{a} ) &\coloneqq - \frac{1}{2} \sum_{ i } \sum_{ j \in \mathcal{ I }_i } \lVert a_i - a_j \rVert,
\end{align}
and proposed using the WLU
\begin{align}
 u\gameref{eq:marden_consensus}_i ( \bm{a} ) &\coloneqq - \sum_{ j \in \mathcal{ I }_i } \lVert a_i - a_j \rVert = \sum_{ j \neq i } \lVert a_i - a_j \rVert \indicator{ ij \in \mathcal{ E } }.
 \end{align}
Since game $\mathcal{G}\game[\label{eq:marden_consensus}] \coloneqq ( \mathcal{ I }, ( \mathcal{ A }_i ), ( u\gameref{eq:marden_consensus}_i ) ) $ is a BSI game with $ w_{ ij }( a_i, a_j ) = - \lVert a_i - a_j \rVert \indicator{ ij \in \mathcal{ E } } $, $ \mathcal{G}\gameref{eq:marden_consensus}$ is confirmed to be an EPG.

\end{example}

\subsection{Identification of Ordinal Potential Games}
\label{ssec:identification_ordinal}
In contrast to EPGs, OPGs have many ordinal potential functions \cite{Monderer1996}.

\begin{theorem}
 \label{eq:ordinal_transform}
Consider the game $ ( \mathcal{ I }, ( \mathcal{A}_i ), ( u_i ) ) $. If there exists a strictly increasing transformation $ f_i \colon \mathbb{R} \to \mathbb{R} $ for every $ i \in \mathcal{ I } $ such that game $ ( \mathcal{ I }, ( \mathcal{A}_i ), ( f_i ( u_i ) ) ) $ is an OPG, the original game $ ( \mathcal{ I }, ( \mathcal{A}_i ), ( u_i ) ) $ is an OPG with the same potential function \cite{Neel2004}.
\end{theorem}

\section{Learning Algorithms}
\label{sec:learning}

A variety of learning algorithms are available to facilitate the convergence of potential games to Nash equilibrium, e.g., myopic best response, fictitious play, reinforcement learning, and spatial adaptive play. Unfortunately, there are no general dynamics that are guaranteed to converge to a Nash equilibrium for a wide class of games \cite{Hart2003}. Since Lasaulce et al.\@ \cite[Sections 5 and 6]{Lasaulce2011} comprehensively summarized these learning algorithms and their sufficient conditions for convergence for various classes of games (including potential games), we present only two frequently used algorithms.

\begin{dfn}
 \textit{Best-response dynamics} refers to the following update rule:  At each step $k$, player $i \in \mathcal{ I }$ \textit{unilaterally} changes his/her strategy from $a_i[k]$ to his/her best response $\bm{a}_{-i}[k]$; in particular,
\begin{align}
 a_i[k + 1] &\in \br_i(\bm{a}_{-i}[k]).
\end{align}
The other players choose the same strategy, i.e., $\bm{a}_{-i}[k + 1] = \bm{a}_{-i}[k]$.
\end{dfn}

Note that while the term ``best-response dynamics'' was introduced by Matsui \cite{Matsui1992}, it has many representations depending on the type of game. We also note that best-response dynamics may converge to sub-optimal Nash equilibria. By contrast, the following spatial adaptive play can converge to the optimal Nash equilibrium. To be precise, it maximizes the potential function with arbitrarily high probability.

\begin{dfn}
Consider a game with a finite number of strategy sets. \textit{Log-linear learning} \cite{Blume1993}, \textit{spatial adaptive play} \cite{Young1998}, and \textit{logit-response dynamics} \cite{Alos-Ferrer2010} refer to the following update rule: At each step $k$, a player $i \in \mathcal{ I }$ unilaterally changes his/her strategy from $a_i[k]$ to $a_i$ with probability $ x_i \in \Delta ( \mathcal{A}_i )$ according to the Boltzmann-Gibbs distribution
\begin{align}
 x_i( a_i \mid \bm{a}_{-i}[k] )
 &= \frac{\exp[{\beta u_i(a_i,\bm{a}_{-i}[k])}]}{\sum_{a_i' \in \mathcal{A}_i}\exp{[\beta u_i(a_i',\bm{a}_{-i}[k])}]}, \label{eq:logit}
\end{align}
where $ \beta\ ( 0 < \beta < \infty ) $ is related to the (inverse) temperature in an analogy to statistical physics. Note that in the limit $\beta\to\infty$, the spatial adaptive play approaches the best-response dynamics.
\end{dfn}

Note that (\ref{eq:logit}) is the solution to the following approximated maximization problem:
\begin{align}
 &\max_{ a_i \in \mathcal{A}_i } u_i ( a_i, \bm{a}_{-i} )
 = \max_{ x_i ( a_i ) } \sum_{ a_i \in \mathcal{A}_i } x_i ( a_i ) \, u_i ( a_i, \bm{a}_{-i}) \nonumber \\
 &\approx \max_{ x_i ( a_i ) } \left[
 \sum_{ a_i \in \mathcal{A}_i } x_i ( a_i ) \, u_i ( a_i, \bm{a}_{-i} )
 - \frac{ 1 }{ \beta } \sum_{ a_i \in \mathcal{A}_i } x_i ( a_i ) \log x_i ( a_i )
 \right], \label{eq:perturbed}
\end{align}
which is called a \textit{perturbed} payoff, where $ \sum_{ a_i \in \mathcal{A}_i } x_i ( a_i ) \allowbreak \log x_i ( a_i ) $ is the entropy function. The derivation of (\ref{eq:logit}) from (\ref{eq:perturbed}) can be found in \cite{Chen2010}.

\begin{theorem}
 [\cite{Blume1993,Young1998}]
In the finite EPG $(\mathcal{ I }, (\mathcal{A}_i), (u_i))$ with potential function $ \phi $, the spatial adaptive play has the unique stationary distribution of strategy profile $ x \in \Delta ( \mathcal{A} ) $, where
\begin{align}
 x(\bm{a}) &= \frac{\exp{[\beta \phi(\bm{a})}]}{\sum_{\bm{a} \in \mathcal{A}}\exp{[\beta \phi(\bm{a})}]},
\end{align}
i.e., it is also the Boltzmann-Gibbs distribution.
\end{theorem}
Further discussion can be found in \cite{Babadi2010,Marden2012revisiting}.

\section{Channel Assignment to Manage Received and Generated Interference Power in TX-RX Pair Model}
\label{ssec:nie}

In the TX-RX pair model shown in Fig.~\ref{fig:system_models}\subref{fig:model2}, Nie and Comaniciu \cite{Nie2006} pointed out that the channel selection game $ \mathcal{G}\gameref{g:u1} $ introduced in Section~\ref{sec:game} was not an EPG. Note that the payoff function of $ \mathcal{G}\gameref{g:u1}$ is the negated sum of received interference from neighboring TXs. To ensure that the channel selection game is an EPG, they considered the channel selection game $\mathcal{G}\game[\label{eq:nie}]$, whose payoff function was the negated sum of the received interference from neighboring TXs, and generated interference to neighboring RXs, i.e.,
\begin{align}
 u\gameref{eq:nie}_i ( \bm{c} ) &\coloneqq -\sum_{j \neq i} ( G_{ij} P_j + G_{ji} P_i ) \indicator{ c_j = c_i }. \label{eq:nie_utility}
\end{align}
Since $ \mathcal{G}\gameref{eq:nie}$ is a BSI game with $w_{ij}(c_i, c_j) = - ( G_{ij}P_j + G_{ji}P_i ) \indicator{ c_j = c_i }$, it is an EPG with potential
\begin{align}
\phi\gameref{eq:nie} (\bm{c})
&= - \sum_{ i } I_i(\bm{c}) = - \sum_{ i } \sum_{ j \neq i } G_{ij} P_j \indicator{ c_j = c_i },
\end{align}
which corresponds to the negated sum of received interference in the entire network. Note that in order to evaluate (\ref{eq:nie_utility}), each pair $i$ needs to estimate or share the values of the generated interference to neighboring RXs, $G_{ji } P_i \indicator{ c_j = c_i }$.

Concurrently with the above, Kauffmann et al.\@ \cite{Kauffmann2007} discussed the following potential function $ \phi\game[\label{eq:kauffmann}] (\bm{ c }) $, which includes RX-specific noise power $ N_i ( c_i ) $, and derived a payoff function using Theorem \ref{th:interactionpotential},
\begin{align}
\phi\gameref{eq:kauffmann} (\bm{ c })
&\coloneqq - \sum_{ i } \sum_{ j \neq i } G_{ij} P_j \indicator{ c_j = c_i } - \sum_{i} N_i( c_i ),  \\
 u{\gameref{eq:kauffmann}}_i ( \bm{ c } ) &= - \sum_{j \neq i} ( G_{ij} P_j + G_{ji} P_i ) \indicator{c_j=c_i} - N_i ( c_i ). \label{eq:kauffmann_utility}
\end{align}

To enable multi-channel allocation, e.g., orthogonal frequency-division multiple access (OFDMA) subcarrier allocation or resource block allocation, La et al.\@ \cite{La2011-11} discussed a modification of $ \mathcal{G} \gameref{eq:nie} $ suitable for multi-channel allocation.

In contrast to unidirectional links assumed in the TX-RX pair model, Uykan and J\"antti \cite{Uykan2014WCOM1,Uykan2014WCOM2} discussed a channel assignment problem for bidirectional links and proposed a joint transmission order and channel assignment algorithm.

\subsection{Joint Transmission Power and Channel Allocation}

Nie et al.\@ \cite{Nie2007} showed that the joint channel selection and power control game with payoff function
\begin{align}
 u{\gameref{eq:nie2}}_i ( \bm{p}, \bm{c} ) &\coloneqq - \sum_{ j \neq i } ( G_{ij} p_j + G_{ji} p_i ) \indicator{c_i=c_j}
 \label{eq:nie2_utility}
\end{align}
is an EPG. Because the best response in $\mathcal{G}\game[\label{eq:nie2}]$ results in the minimum transmission power level, Bloem et al.\@ \cite{Bloem2007} proposed adding terms $\alpha \log (1 + G_{ii}p_i) + \beta / p_i$ to (\ref{eq:nie2_utility}) to account for the achievable data rate and consumed power. Note that these terms are self-motivated functions, and the game with the modified payoff function is still an EPG.

As another type of joint assignment, a preliminary beamform pattern setting followed by channel allocation was discussed in \cite{Zeydan2008-10}.

\subsection{Primary-secondary Scenario and Heterogeneous Networks}

To manage interference in primary-secondary systems, Bloem et al.\@ \cite{Bloem2007} proposed adding terms related to the received and generated interferences from and to the primary user. They also proposed adding cost terms related to payoff function (\ref{eq:nie2_utility}). In particular, they discussed a Stackelberg game \cite{Myerson1991}, where the primary user was the leader and the secondary users were followers. Giupponi and Ibars discussed overlay cognitive networks \cite{Giupponi2009-23eurasip} and heterogeneous OFDMA networks \cite{Giupponi2010-25}. Mustika et al.\@ \cite{Mustika2010-0potential} took a similar approach to prioritize users.

Uplinks of heterogeneous OFDMA cellular systems with femtocells were discussed in \cite{Mustika2011-18}, whereas downlinks of OFDMA cellular systems, where each BS transmits to several mobile stations, were discussed in \cite{La2012a-0,La2012-6}. OFDMA relay networks were considered in  \cite{Liang2012-9}. Further discussion can be found in \cite{Huang2015VT}. Joint BS/AP selection and channel selection problems were discussed in \cite{Dai2015JSAC}.

\section{Channel Assignment to Enhance SINR and Throughput in TX-RX Pair Model}
\label{sec:iSINR}

In the TX-RX pair model shown in Fig.~\ref{fig:system_models}\subref{fig:model2}, the signal-to-interference-plus-noise ratio (SINR) at RX $ i $ is given by
\begin{align}
  \frac{ G_{ii} P_i }{N_i + I_i (\bm{c}) }
 = \frac{ G_{ii} P_i }{N_i + \sum_{ j \neq i } G_{ij} P_j \indicator{ c_j = c_i }} &\eqqcolon \mathrm{SINR}_i(\bm{c}). \label{eq:sinr_buzzi}
\end{align}
Menon et al.\@ \cite{Menon2009interference} pointed out that there may be no Nash equilibrium in the channel selection game $ ( \mathcal{ I }, ( \mathcal{ C }_i ), (\mathrm{SINR}_i) ) $.

Instead, they proposed using the sum of the inverse SINR, defined by
\begin{align}
 u\gameref{eq:buzzi}_i (\bm{c}) &\coloneqq -  \frac{ 1 }{ \mathrm{SINR}_i(\bm{c}) } - \sum_{j \neq i} \frac{G_{ji} P_i}{G_{jj} P_j} \indicator{c_j = c_i}, \label{eq:buzzi_utility}
\end{align}
as the  payoff function. Similar to $ \mathcal{G} \gameref{eq:nie}$, $\mathcal{G}\game[\label{eq:buzzi}] \coloneqq ( \mathcal{ I }, ( \mathcal{ C }_i ), ( u\gameref{eq:buzzi}_i ) )$ is a BSI game with $ w_{ij}(c_i, c_j) = - [( G_{ij} P_j / G_{ii} P_i ) +  ( G_{ji} P_i / G_{jj} P_j ) ] \indicator{ c_j = c_i } $. Thus, $\mathcal{G}\ref{eq:buzzi}$ is an EPG with potential
\begin{align}
 \phi{\gameref{eq:buzzi}} ( \bm{c} ) &= - \sum_{i} \frac{ 1 }{ \mathrm{SINR}_i(\bm{c}) },
\end{align}
i.e., the sum of the inverse SINR in the network.

Note that the above expression is a single carrier version of orthogonal channel selection. Menon et al.\@ \cite{Menon2009interference} discussed a waveform adaptation version of $ \mathcal{G} \gameref{eq:buzzi} $ that can be applied to codeword selection in non-orthogonal code division multiple access (CDMA), and Buzzi et al.\@ \cite{Buzzi2013} further discussed waveform adaptation. Buzzi et al.\@ \cite{Buzzi2012} also discussed an OFDMA subcarrier allocation version of $ \mathcal{G} \gameref{eq:buzzi}$.
Cai et al.\@ \cite{Cai2015WPC} discussed joint transmission power and channel assignment utilizing the payoff function (\ref{eq:buzzi_utility}) of $ \mathcal{G} \gameref{eq:buzzi}$.

G\'{a}llego et al.\@ \cite{Gallego2012} proposed using the network throughput of joint power and channel assignment,
\begin{align}
 \sum_i \indicator{ \mathrm{ SINR }_i(\bm{p},\bm{c}) \geq \varGamma } \, B_{ c_i } \log \left( 1 + \mathrm{ SINR }_i(\bm{p},\bm{c}) \right),
 \label{eq:gallego}
\end{align}
as potential, where $ B_{ c_i } $ is the bandwidth of channel $ c_i $, and $ \varGamma $ is the required SINR. It may have been difficult to derive a simple payoff function, and they thus proposed the WLU (\ref{eq:WLU2}) of (\ref{eq:gallego}).

\section{Channel Assignment to Manage the Number of Interference Signals in TX-RX Pair Model}
\label{ssec:number_interference}

Yu et al.\@ \cite{Yu2010-52} and Chen et al.\@ \cite{Chen2011-19} considered sensor networks where each RX (sink) receives messages from multiple TXs (sensors). They proved that a channel selection that minimizes the number of received and generated interference signals is an EPG, where the potential is the number of total interference signals. Note that the average number of retries is approximately proportional to the number of received interference signals when the probability that the messages are transmitted is very small, as in sensor networks.

A simpler and related form of (\ref{eq:nie_utility}) is detailed in the following discussion. To reduce the information exchange required to evaluate (\ref{eq:nie_utility}), Yamamoto et al.\@ \cite{Yamamoto2010} proposed using the number of received and generated interference sources as the payoff function, where the received interference power is greater than a given threshold $T$, i.e.,
\begin{align}
 u{\gameref{eq:yamamoto}}_i ( \bm{ c } ) &\coloneqq -\sum_{j \neq i} \left( \indicator{G_{ij}P_j > T} + \indicator{G_{ji}P_i > T} \right) \indicator{c_j = c_i}.
 \label{eq:yamamoto_utility}
\end{align}
This model is sometimes referred to as a ``binary'' interference model \cite{Maheshwari2008} in comparison with a ``physical'' interference model. Because $\mathcal{G}\game[\label{eq:yamamoto}] \coloneqq ( \mathcal{ I }, ( \mathcal{ C }_i ), (u\gameref{eq:yamamoto}_i) ) $ is a BSI game with $ w_{ ij } ( c_i, c_j ) = - ( \indicator{G_{ij}P_j > T} + \indicator{G_{ji}P_i > T} ) \indicator{c_j = c_i} $, $\mathcal{G}\gameref{eq:yamamoto}$ is an EPG. When we consider a directed graph, where edges between TX $ j $ and RX $ i $ indicate $ G_{ij} P_j > T$, we denote TX $ i $'s neighboring RXs by $ \mathcal{ R }_i \coloneqq \{\, j \in \mathcal{ I } \mid j \neq i\ \text{and}\ ji \in \mathcal{ E } \,\} $,  and RX $ i $'s neighboring TXs by $ \mathcal{ T }_i \coloneqq \{\, j \in \mathcal{ I } \mid j \neq i\ \text{and}\ ij \in \mathcal{ E } \,\} $. Using these expressions, (\ref{eq:yamamoto_utility}) can be rewritten to
\begin{align}
 u{\gameref{eq:yamamoto}}_i ( \bm{ c } ) &=
 - \sum_{ j \neq i } \left( \indicator{ij \in \mathcal{ E } } + \indicator{ ji \in \mathcal{ E } } \right) \indicator{c_j = c_i} \nonumber \\
 &= - \sum_{ j \in \mathcal{ T }_i } \indicator{ c_j = c_i }
 - \sum_{ j \in \mathcal{ R }_i } \indicator{ c_j = c_i }.
\end{align}
Yang et al.\@ \cite{Yang2012} discussed a multi-channel version of $ \mathcal{G} \gameref{eq:yamamoto}$.

\section{Channel Assignment to Manage Received Interference Power in TX Network Model}
\label{sec:neel}

\subsection{Identical Transmission Power Levels}

In Section \ref{ssec:nie}, channel allocation games in the TX-RX pair model shown in Fig.~\ref{fig:system_models}\subref{fig:model2} are discussed. Neel et al.\@ \cite{Neel2006-55performance,Neel2007-14interference} considered a different channel allocation game typically applied to channel allocation for APs in the wireless local area networks (WLANs) shown in Fig.~\ref{fig:system_models}\subref{fig:model3}, where each TX $i \in \mathcal{ I }$ selects a channel $c_i \in \mathcal{C}_i$ to minimize the interference from other TXs, i.e.,
\begin{align}
 u{\gameref{eq:neel}}_i (\bm{c}) &\coloneqq - I_i(\bm{c}) \coloneqq - \sum_{ j\neq i } G_{ij} P \indicator{c_i = c_j}, \label{eq:neel_utility}
\end{align}
where $P$ is the common transmission power level for every TX. Note that $ G_{ij} = G_{ji} $ in this scenario, whereas $ G_{ij} \neq G_{ji} $ in the TX-RX pair model shown in Fig.~\ref{fig:system_models}\subref{fig:model2}. Moreover, note that interference from stations other than the TXs is not taken into account in the payoff function. In addition to the TX network model, channel selection can be applied to the canonical network model shown in Fig.~\ref{fig:system_models}\subref{fig:model4} \cite{Babadi2010}.

Because $\mathcal{G}\game[\label{eq:neel}]$ is a BSI game where $w_{ij}(c_i, c_j) = -G_{ij}P\indicator{c_i = c_j}$, it is an EPG with potential
\begin{align}
 \phi{\gameref{eq:neel}} (\bm{c}) &= -\frac{1}{2} \sum_{\mathclap{i}} I_i(\bm{c}),
\end{align}
which corresponds to the aggregated interference power among TXs. Neel et al.\@ pointed out that other symmetric interference functions, e.g., $ \max \{ B-\lvert c_i - c_j \rvert, 0 \}/B $, where $B$ is the common bandwidth for every channel, can be used instead of $\indicator{c_i = c_j}$ in (\ref{eq:neel_utility}).

Kauffmann et al.\@ \cite{Kauffmann2007} discussed essentially the same problem. However, they considered player-specific noise, and derived (\ref{eq:neel_utility}) by substituting $ G_{ij} = G_{ji} $ and $ P_i = P_j = P $ into (\ref{eq:kauffmann_utility}).

Compared with the payoff function (\ref{eq:nie_utility}), (\ref{eq:neel_utility}) can be evaluated with only local information available at each TX; however, the transmission power levels of all TXs need to be identical. We further discuss this requirement in Section \ref{ssec:non-identical}.

Liu and Wu \cite{Liu2008} reformulated the game represented by (\ref{eq:neel_utility}) as a CG by introducing virtual resources. Further discussion can be found in \cite{Li2012distributed}.

\subsection{Non-identical Transmission Power Levels}
\label{ssec:non-identical}

To avoid the requirement of identical transmission power levels in (\ref{eq:neel_utility}), Neel \cite{Neel2007-5synthetic} proposed using the product of (constant) transmission power level $ P_i $ and interference $ I_i(\bm{c}) $ as the payoff function, i.e.,
\begin{align}
 u{\gameref{eq:babadi}}_i (\bm{c}) &\coloneqq - P_i I_i(\bm{c}) = - P_i \sum_{j \neq i} G_{ij} P_j \indicator{c_j = c_i}. \label{eq:babadi_utility}
\end{align}
Because $\mathcal{G}\game[\label{eq:babadi}]$ is a BSI game with $w_{ij} = -P_i G_{ij} P_j \indicator{c_j = c_i}$, $\mathcal{G}\gameref{eq:babadi}$ is an EPG with
\begin{align}
 \phi{\gameref{eq:babadi}} (\bm{c}) &= - \frac{1}{2} \sum_{i} P_i \sum_{j \neq i} G_{ij} P_j \indicator{c_j = c_i}. \label{eq:babadi_V}
\end{align}
Note that this form of payoff functions was provided by Menon et al.\@ \cite{Menon2009game} in the context of waveform adaptations.
This game under frequency-selective channels was discussed by Wu et al.\@ \cite{Wu2012}.

The relationship between (\ref{eq:babadi_utility}) and its exact potential function (\ref{eq:babadi_V}) implies that the game $ \mathcal{G}\game[\label{eq:babadi_weighted}] $ with payoff function
\begin{align}
 u{\gameref{eq:babadi_weighted}}_i (\bm{c}) &\coloneqq -I_i(\bm{c}) = - \sum_{j\neq i} G_{ij} P_j \indicator{c_j = c_i}
\end{align}
is a WPG with potential function $ \phi{\gameref{eq:babadi}} (\bm{c}) $ and $\alpha_i = 1/P_i$ in (\ref{eq:wpg}), i.e., the identical transmission power level required in (\ref{eq:neel_utility}) is not necessarily required for the game to have the FIP. This was made clear by Bahramian et al.\@ \cite{Bahramian2008-9} and Babadi et al.\@ \cite{Babadi2010}.

As extensions, in \cite{Wang2012}, the interference management game $\mathcal{G}\game[\label{eq:Wang2012}]$ on graph structures with the following payoff function was discussed:
\begin{align}
 u{\gameref{eq:Wang2012}}_i ( \bm{ c } ) &\coloneqq - P_i \sum_{ i \in \mathcal{ I }_i } G_{ij} P_j \indicator{ c_j = c_i } \nonumber \\ &= - P_i \sum_{ i } G_{ij} P_j \indicator{ c_j = c_i } \indicator{ ji \in \mathcal{ E } }.
\end{align}
\cite{Wu2013,Zheng2014VT} proposed using the expected value of interference in order to manage fluctuating interference.
Zheng \cite{Zheng2015WCOML} treated dynamical on-off according to traffic variations in $ \mathcal{G}\gameref{eq:babadi_weighted} $.

\section{Channel Assignment to Enhance SINR and Capacity in TX Network Model}
\label{eq:neel_capacity}

Menon et al.\@ \cite{Menon2009game} showed that a waveform adaptation game where the payoff function is the SINR or the mean-squared error at the RX is an OPG. Chen and Huang \cite{Chen2013} showed that a channel allocation game in the TX network model shown in Fig.~\ref{fig:system_models}\subref{fig:model3}, or in the canonical network model shown in Fig.~\ref{fig:system_models}\subref{fig:model4}, where the payoff function is the SINR or a Shannon capacity, is an OPG. Here, we provide a derivation in the form of channel allocation according to the derivation provided in \cite{Menon2009game}. A channel selection game $\mathcal{G}\game[\label{eq:PIN}]$ with payoff function
\begin{align}
 u{\gameref{eq:PIN}}_i (\bm{c}) &\coloneqq -P_i [ N_i(c_i) + I_i(\bm{c}) ]
 \label{eq:PIN_utility}
\end{align}
is an EPG with potential
\begin{align}
 \phi{\gameref{eq:PIN}} (\bm{c}) &= - \sum_{i} P_i N_i(c_i) - \frac{1}{2}\sum_{i} P_i I_i( \bm{ c } ).
\end{align}
Because $P_i$ is a constant in (\ref{eq:PIN_utility}), by Theorem \ref{eq:ordinal_transform}, $\mathcal{G}\game[\label{eq:SINR_chen}]$ with payoff
\begin{align}
 u{\gameref{eq:SINR_chen}}_i (\bm{c}) &\coloneqq \frac{- G_{ ii } { P_i }^2}{ u{\gameref{eq:PIN}}_i (\bm{ c }) }
 = \frac{G_{ii} P_i}{ N_i(c_i) + I_i(\bm{c}) }
\end{align}
is an OPG with potential $\phi{\gameref{eq:PIN}} (\bm{c})$. As a result, once again using Theorem \ref{eq:ordinal_transform}, $\mathcal{G} \game[\label{eq:capacity_chen}]$ with payoff
\begin{align}
 u{\gameref{eq:capacity_chen}}_i (\bm{c}) &\coloneqq B \log \left( 1+ u{\gameref{eq:SINR_chen}}_i ( \bm{ c } ) \right) \nonumber \\
 & = B \log \left( 1 + \frac{G_{ii} P_i}{ N_i(c_i) + I_i(\bm{c}) } \right)
 \label{eq:capacity_chen_utility}
\end{align}
is an OPG with potential $\phi{\gameref{eq:PIN}} (\bm{c})$.
Xu et al.\@ \cite{Xu2015arxiv_database} further discuss $ \mathcal{G} \gameref{eq:capacity_chen} $, where the active TX set can be stochastically changed.

A quite relevant discussion was conducted by Song et al.\@ \cite{Song2008joint}. They discussed a joint transmission power and channel assignment game $\mathcal{G} \game[\label{eq:song_capacity}] $ to maximize throughput:
\begin{align}
 u{\gameref{eq:song_capacity}}_i (\bm{p}, \bm{c}) &\coloneqq R \left( 1 + \frac{G_{ii} p_i}{ N_i(c_i) + I_i(\bm{p},\bm{c}) } \right),
\end{align}
where $ R : \mathbb{R} \to \mathbb{R} $ represents throughput depending on SINR. They pointed out that since each user would set the maximum transmission power at a Nash equilibrium, $\mathcal{G} \gameref{eq:song_capacity} $ is equivalent to the channel selection game $ \mathcal{G} \gameref{eq:neel}$.
Further discussion on joint transmission power and channel assignment can be found in \cite{Maghsudi2014VT}.

\section{Channel Assignment to Manage the Number of Interference Signals in Interference Graph}
\label{sec:min_number}

For the interference graph $ ( \mathcal{ I }, \mathcal{ E } ) $ shown in Fig.~\ref{fig:system_models}\subref{fig:model5}, Xu et al.\@ \cite{Xu2012-42STSP} proposed using the number of neighbors that select the same channel as the payoff function, i.e.,
\begin{align}
 u{\gameref{eq:xustsp1}}_i ( \bm{ c } ) &\coloneqq - \sum_{ j \in \mathcal{ I }_i } \indicator{ c_j = c_i }.
 \label{eq:xustsp1_utility}
\end{align}
We would like to point out that (\ref{eq:xustsp1_utility}) can be reformulated to
\begin{align}
 u{\gameref{eq:xustsp1}}_i ( \bm{ c } ) &= - \sum_{ j \neq i } \indicator{ c_j = c_i } \indicator{ ij \in \mathcal{ E } },
\end{align}
i.e., $\mathcal{G} \game[\label{eq:xustsp1}] $ is a BSI game with $ w_{ ij }( c_i, c_j ) = - \indicator{ c_j = c_i } \indicator{ ij \in \mathcal{ E } } $. Thus $ \mathcal{G} \gameref{eq:xustsp1} $ is an EPG. Note that this is a special case of singleton CGs on graphs discussed in Section \ref{ssec:aloha}.

As variations of $ \mathcal{G} \gameref{eq:xustsp1} $, Xu et al.\@ \cite{Xu2013JCOM} discussed the impact of partially overlapped channels. Yuan et al.\@ \cite{Yuan2013} discussed the variable-bandwidth channel allocation problem. Zheng et al.\@ \cite{Zheng2015VT} took into account stochastic channel access according to the carrier sense multiple access (CSMA) protocol. Xu et al.\@ \cite{Xu2015arxiv_centralized} discussed a multi-channel version of $ \mathcal{G} \gameref{eq:xustsp1}$.

Liu et al.\@ \cite{Liu2013WCNC} discussed a common control channel assignment problem for cognitive radios, and proposed using $ \sum_{ j \neq i } \indicator{ c_j = c_i } $ for the payoff function so that every player chooses the same channel. This game is similar to the consensus game $ \mathcal{G}\gameref{eq:marden_consensus}$.

\section{Channel Assignment to Enhance Throughput in Collision Channels}
\label{sec:ca_aloha}

Channels can be viewed as common resources in the congestion model introduced in Section \ref{sssec:congestion_game}. In general, throughput when using a channel depends only on the number of stations that select the relevant channel. A CG formulation is thus frequently used for channel selection problems. Altman et al.\@ \cite{Altman2009} formulated a multi-channel selection game in a single collision domain as a CG. Based on a CG formulation, channel selections by secondary stations were discussed in \cite{Xu2012-43TW,Jiao2014}. A channel selection problem in multiple collision domains was discussed in \cite{Xu2013TW}. Iellamo et al.\@ \cite{Iellamo2013} used numerically evaluated successful access probabilities depending on the number of stations in CSMA/CA as payoff functions.

Here, we discuss channel selection problems in interference graph $ ( \mathcal{ I }, \mathcal{ E } ) $, where each node $ i \in \mathcal{ I } $ attempts to adjust its channel $ c_i $ to maximize its successful access probability or throughput.

\subsection{Slotted ALOHA}
\label{ssec:aloha}

Consider collision channels shared using slotted ALOHA. Each node $ i $ adjusts its channel to avoid simultaneous transmissions on the same channel because these result in collisions. In this case, when one node exclusively chooses a channel, the node can transmit without collisions. Thus, the following payoff function captures the benefit of nodes:
\begin{align}
 u{\gameref{eq:thomas}}_i ( \bm{ c } ) &\coloneqq
 \begin{cases}
  1 & \text{if}\ \lvert \mathcal{ I }_{i}^{c_i} \rvert = 0, \\
  0 & \text{otherwise}.
 \end{cases}
\end{align}
$\mathcal{G} \game[\label{eq:thomas}] $ is a singleton CG on graphs, and Thomas et al.\@ \cite{Thomas2007} showed that $ \mathcal{G} \gameref{eq:thomas} $ is an OPG\footnote{There is another simple proof of this based on the fact that $ \mathcal{G} \gameref{eq:thomas} $ is equivalent to $ \mathcal{G} \gameref{eq:aloha_chen} $ when setting $ X_i = 1 $ for every $ i $.}.

Consider that each node has a transmission probability $ X_i $ ($ 0 < X_i < 1 $). Chen and Huang \cite{Chen2013JSAC_distributed} proposed using the logarithm of successful access probability,
\begin{align}
  u{\gameref{eq:chen_aloha_jsac}}_i ( \bm{ c } ) &\coloneqq \log \left[
 X_i \textstyle\prod_{ j \in \mathcal{ I }_{ i }^{ c_i } } ( 1 - X_j )
 \right]
\end{align}
and proved that $\mathcal{G}\game[\label{eq:chen_aloha_jsac}]$ is a WPG. Here, we provide a different proof. When we consider
\begin{align}
 &u{\gameref{eq:chen_aloha_jsac2}}_i ( \bm{ c } ) \coloneqq - \log (1-X_i) \cdot u{\gameref{eq:chen_aloha_jsac}}_i ( \bm{ c } )  \\
 &= - \log (1 - X_i)  \log \left[
 X_i \textstyle\prod_{ j \neq i } ( 1 - X_j )^{\indicator{ c_j = c_i} \indicator{ ij \in \mathcal{ E } } }
 \right] \nonumber \\
 &= - \log (1 - X_i)  \log (X_i)   \nonumber\\
 &\hphantom{ = {} } - \log (1 - X_i)  \textstyle\sum_{ j \neq i } \indicator{ c_j = c_i } \indicator{ ij \in \mathcal{ E } } \log ( 1 - X_j ) , \nonumber
\end{align}
$\mathcal{G} \game[\label{eq:chen_aloha_jsac2}] $ is a BSI game with $w_{ij}(c_i, c_j) = - \log(1 - X_i) \log(1 - X_j) \indicator{ c_j = c_i } \indicator{ ij \in \mathcal{ E } } $. Thus, $ \mathcal{G} \gameref{eq:chen_aloha_jsac} $ is a WPG and, by Theorem \ref{eq:ordinal_transform}, $\mathcal{G} \game[\label{eq:aloha_chen}]$ with payoff
\begin{align}
 u{\gameref{eq:aloha_chen}}_i (\bm{ c }) &\coloneqq X_i \textstyle\prod_{ j \in \mathcal{ I }_{ i }^{ c_i }( \bm{ c } ) } (1-X_j).
\end{align}
is an OPG.
Chen and Huang \cite{Chen2015spatial} further discussed $ \mathcal{G} \gameref{eq:aloha_chen} $ with player-specific constants and proved that the game is an OPG.

Before concluding this section, we would like to point out the relationship between $ \mathcal{G}\gameref{eq:chen_aloha_jsac}$ and CGs. When we assume an identical transmission probability $ X_i = X $ for every $ i $, we get
\begin{align}
 u{ \gameref{eq:chen_aloha_jsac} }_i ( \bm{ c } )
 &= \log (X) + \log(1-X) \textstyle\sum_{ j \neq i } \indicator{  c_j = c_i } \indicator{ ij \in \mathcal{ E } },
\end{align}
i.e., $ \mathcal{G} \gameref{eq:chen_aloha_jsac}$ is a CG on graphs.

\subsection{Random Backoff}

Let the backoff time of player $ i $ be denoted by $ \lambda_i \in [1, \lambda_\mathrm{max}] $, where $ \lambda_\mathrm{max} $ represents the backoff window size. The probability to acquire channel access is given by
\begin{align}
 u{\gameref{eq:chen_random_backoff1}}_i ( \bm{ c } ) &\coloneqq \Pr\left\{ \lambda_i < \min_{ j \in \mathcal{ I }_i^{ c_i } } \{ \lambda_j \} \right\} \nonumber \\
 &= \sum_{\lambda = 1}^{\lambda_\mathrm{max}} \frac{1}{\lambda_\mathrm{max}} \left(\frac{\lambda_\mathrm{max}-\lambda}{\lambda_\mathrm{max}}\right)^{ \sum_{ j \neq i } \indicator{ c_j = c_i } }.
\end{align}
$\mathcal{G} \game[\label{eq:chen_random_backoff1}]$ is a singleton CG, and thus is an EPG. Furthermore, $\mathcal{G} \game[\label{eq:chen_random_backoff2}] $ with
\begin{align}
 u{\gameref{eq:chen_random_backoff2}}_i (\bm{c}) \coloneqq
\lim_{ \lambda_\mathrm{max} \to \infty } u{\gameref{eq:chen_random_backoff1}}_i (\bm{c}) = \frac{ 1 }{ 1 + \sum_{ j \neq i } \indicator{ c_j = c_i } }
\end{align}
is also a singleton CG.

Chen and Huang \cite{Chen2013} showed that $ \mathcal{G} \gameref{eq:chen_random_backoff1} $ with player-specific constants is an OPG. They \cite{Chen2015spatial} further discussed $ \mathcal{G} \gameref{eq:chen_random_backoff1} $ and $ \mathcal{G} \gameref{eq:chen_random_backoff2} $ on graphs with player-specific constants, and proved that these are OPGs according to the proof provided in \cite{Mavronicolas2007}. Xu et al.\@ \cite{Xu2015arxiv} further discussed the game under fading channels.

Chen and Huang \cite{Chen2015spatial} generalized $ \mathcal{G} \gameref{eq:chen_random_backoff2} $ to $\mathcal{G} \game[\label{g:chen_weight}]$, whose payoff function is a generalized throughput 
\begin{align}
 u{\gameref{g:chen_weight}}_i ( \bm{ c } ) \coloneqq \frac{ w_i }{ \sum_{ j } w_j },
\end{align}
where $ w_i\ (> 0) $ represents the channel-sharing weight for player $ i $. Du et al.\@ \cite{Du2015WCOM} further discussed this kind of game.

For the TX-RX pair model shown in Fig.~\ref{fig:system_models}\subref{fig:model2}, Canales and G\'{a}llego \cite{Canales2010} proposed using the following network throughput as a result of joint transmission power and channel assignment as potential:
\begin{align}
\sum_{ i } \frac{ B_{ c_i } }{ 1 + \sum_{ j \neq i } \indicator{ G_{ij} p_j > T } \indicator{c_j = c_i} } \log_2 \left( 1 + \frac{G_{ii}p_i}{ N }\right),
 \label{eq:canales}
\end{align}
where $ G_{ ij } \neq G_{ ji } $, $ B_{ c_i } $ is the bandwidth of channel $ c_i $, and $ T $ is the power threshold of interference. Since (\ref{eq:canales}) is too complex, it may be difficult to derive simple payoff functions. Thus, Canales and G\'{a}llego proposed using payoff functions of the form of a WLU (\ref{eq:WLU2}). Note that the evaluation of the WLU of (\ref{eq:canales}) requires the impact of joint assignment on the throughput of neighboring nodes.

\section{Transmission Probability Adjustment for the Multiple-access Collision Channel (Slotted ALOHA)}
\label{sec:alpha}

Consider a collision channel shared using slotted ALOHA. Each node $ i $ adjusts its transmission probability $ x_i \in [0, 1] $ to maximize the following successful access probability (minus the cost):
\begin{align}
 u{\gameref{eq:SALOHA}}_i ( x_i, \bm{ x }_{ -i } ) &\coloneqq
 x_i \textstyle\prod_{ j \neq i } ( 1 - x_j ) - \cost_i ( x_i ).
 \label{eq:SALOHA_utility}
\end{align}
This is a well-known payoff function. Further discussion can be found in \cite{Su2011}. Because (\ref{eq:SALOHA_utility}) satisfies (\ref{eq:twice_difference}), $ \mathcal{G}\gameref{eq:SALOHA} \coloneqq ( \mathcal{ I }, ([0,1]), ( u\gameref{eq:SALOHA}_i ) ) $ is an EPG with potential
\begin{align}
 \phi{\gameref{eq:SALOHA}} ( \bm{ x } ) &= - \textstyle\prod_{ i } ( 1 - x_i ) - \textstyle\sum_{ i } \cost_i ( x_i ).
\end{align}
Candogan et al.\@ \cite{Candogan2009} showed that $ \mathcal{G}\game[\label{eq:SALOHA}] $ in stochastic channel model, where each player adjusts his/her transmission probability based on the channel state, is a WPG.
Cohen et al.\@ \cite{Cohen2013} discussed a multi-channel version of $\mathcal{G}\gameref{eq:SALOHA}$.
They also discussed $ \mathcal{G} \gameref{eq:SALOHA}$ on graphs \cite{Cohen2015WIOPT}.

For this kind of transmission probability adjustment to satisfy $ \sum x_i < 1 $, the cost function $ \cost_i ( \bm{x} ) = \indicator{ \sum_i x_i > 1 } $ needs to be used \cite{Derakhshani2014}. Because this cost function is a coordination function, a game with this cost function is still an EPG.

\section{Transmission Power Assignment to Enhance Throughput in Multiple-access Channel}
\label{sec:tpc}

Here, we discuss power control problems in multiple-access channels, as shown in Fig.~\ref{fig:system_models}\subref{fig:model1}, where each TX attempts to adjust its transmission power level to maximize its throughput. For a summary of transmission power control, we refer to \cite{Chiang2008}. Note that Saraydar et al.\@ \cite{Saraydar2002} applied a game-theoretic approach to an uplink transmission power control problem in a CDMA system. The relation between potential games and transmission power control to achieve target SINR or target throughput has been discussed in \cite{Neel2004}.

Alpcan et al.\@ \cite{Alpcan2002} formulated uplink transmission power control in a single-cell CDMA as the game $ \mathcal{G}\game[\label{eq:cdma}] \coloneqq (\mathcal{ I }, (\mathcal{P}_i), (u\gameref{eq:cdma}_i))$, where $\mathcal{P}_i \coloneqq \{\, p_i \mid 0 < P_{i, \mathrm{min}} \leq p_i \leq P_{i, \mathrm{max}} \,\} $, $P_{i, \mathrm{min}}$ is the minimum transmission power, and $P_{i,\mathrm{max}}$ is the maximum transmission power. In this game, each TX $ i \in \mathcal{ I } $ adjusts its transmission power $p_i \in \mathcal{P}_i$ to maximize its data rate (throughput), which is assumed to be proportional to the Shannon capacity, minus the cost of transmission power, i.e.,
\begin{align}
 u{\gameref{eq:cdma}}_i (p_i,\bm{p}_{-i}) &\coloneqq \log\left(1+\mathit{S} \frac{ G_{i}p_i }{ N + \sum_{j \neq i} G_{j} p_j }\right) - \alpha_i p_i, \label{eq:cdma_utility}
\end{align}
where $\mathit{S}\ (> 1)$ is the spreading gain and $\alpha_i$ is a positive real number. The cost function $ - \alpha_i p_i$ is used to avoid an inefficient Nash equilibrium, where all TXs choose the maximum transmission power. All TXs choose this power because $\br(\bm{p}_{-i})$ is the maximum transmission power for every TX when the cost function is not used \cite{Saraydar2002,Alpcan2002}.

Alpcan et al.\@ \cite{Alpcan2002} proved the existence and uniqueness of a Nash equilibrium in the game, and Neel \cite[\S5.8.3.1]{Neel2006-216analysis} showed that this game is not an EPG because (\ref{eq:twice_difference}) does not hold. Note that Neel et al.\@ \cite{Neel2002game} was the first to apply the potential game approach to this type of power control.

Instead of $ \mathcal{G}\gameref{eq:cdma}$, Fattahi and Paganini \cite{Fattahi2005} proposed setting $\mathit{S} = 1$ in $ \mathcal{G}\gameref{eq:cdma}$, i.e.,
\begin{align}
 &u{\gameref{eq:fattahi}}_i( \bm{ p } ) \coloneqq \log \left(
 1 + \frac{ G_i p_i }{ N + \sum_{j \neq i} G_j p_j }
 \right)
 - \cost_i (p_i) \label{eq:fattahi_utility}\\
 &= \log \left( N + \textstyle\sum_i G_i p_i \right)  - {\log \left( N + \textstyle\sum_{j \neq i} G_j p_j \right)} - \cost_i (p_i), \nonumber
\end{align}
where $\cost_i \colon \mathcal{P}_i \to \mathbb{R}$ is a non-decreasing convex cost function. Since $ u\gameref{eq:fattahi}_i ( \bm{ p } ) $ is a linear combination of a coordination function $\log ( \sum_i G_i p_i + \sigma^2)  $, a dummy function $ \log ( \sum_{j \neq i} G_j p_j + \sigma^2 )$, and a self-motivated function $\cost_i (p_i)$, $ \mathcal{G}\gameref{eq:fattahi} \coloneqq (\mathcal{ I }, ( \mathcal{ P }_i ) , ( u\gameref{eq:fattahi}_i ) ) $ is an EPG with potential
\begin{align}
 \phi\gameref{eq:fattahi}( \bm{ p } ) &= \log \left(
 N + \textstyle\sum_{i} G_i p_i
 \right)
 - \textstyle\sum_i \cost_i (p_i).
\end{align}

Because $ \phi\gameref{eq:fattahi}( \bm{ p } ) $ is continuously differentiable and strictly concave, by Theorem \ref{th:unique}, there is a unique maximizer for the potential, and best-response dynamics converge to a unique Nash equilibrium, which is the maximizer of the potential on strategy space $\prod_i \mathcal{P}_i$. Kenan et al.\@ \cite{Kenan2011} discussed $\mathcal{G}\game[\label{eq:fattahi}]$ over time-varying channels.

Neel \cite[\S5.8.3.1]{Neel2006-216analysis} approximated (\ref{eq:cdma_utility}) by
\begin{align}
 u\gameref{eq:neel_cap}_i ( \bm{ p } ) &\coloneqq \log\left(\mathit{S} \frac{ G_{i} p_i }{ N + \sum_{j \neq i} G_{j} p_j }\right) -  \cost_i(p_i), 
\end{align}
and showed that $\mathcal{G}\game[\label{eq:neel_cap}] \coloneqq ( \mathcal{ I }, ( \mathcal{ P }_i ), ( u\gameref{eq:neel_cap}_i ) ) $ is an EPG with potential $ \phi\gameref{eq:neel_cap}(\bm{p}) = \sum_i(\log p_i - \cost_i(p_i))$. Candogan et al.\@ \cite{Candogan2010} applied $ \mathcal{G} \gameref{eq:neel_cap} $ to multi-cell CDMA systems, and verified that the modified game is an EPG with a unique Nash equilibrium by applying Theorem \ref{th:unique}. A more general form of payoff functions of SINR was discussed in \cite{Ghosh2015WIOPT}.

\subsection{Multi-channel Systems}

A transmission power control game $ \mathcal{G}\gameref{eq:fattahi} $ with multiple channels was discussed in \cite{Jing2009,He2010,Perlaza2009}. Let the set of channels be denoted by $\mathcal{C}$. Each TX $i \in \mathcal{ I }$ transmits through a subset of $ \mathcal{ C } $ to maximize the aggregated capacity
\begin{align}
 \sum_{c \in \mathcal{C}}  \log \left(1 + \frac{G_{i, c} p_{i, c}}{ N_c + \sum_{j\neq i} G_{j, c} p_{j, c} } \right)
 \label{eq:Perlaza}
\end{align}
by adjusting the transmission power vector $ ( p_{ i, 1 }, \dots, p_{ i, \lvert \mathcal{ C } \rvert } ) $. This game is an EPG with potential $ \textstyle\sum_{c \in \mathcal{C}}  \log \left({ N_c + \textstyle\sum_{i} G_{i, c} p_{i, c} } \right) $. Mertikopoulos et al.\@ \cite{Mertikopoulos2012} further discussed the game under fading channels. Note that multi-channel transmission power assignment problems can be seen as joint transmission power and channel assignment problems introduced in Section \ref{eq:neel_capacity} because a zero transmission power level means that the relevant channel has not been assigned \cite{Perlaza2013}.

Note that \cite{Perlaza2009} also discussed BS selection, and further discussion can be found in \cite{Hong2013}. The joint transmission power and bandwidth assignment problem for relay networks was discussed in \cite{Al-Tous2013}.
Primary-secondary scenario \cite{DelRe2009WPC} and heterogeneous network scenario \cite{Ling2013,Wang2014,Zhang2014} were also discussed.

\subsection{Precoding}
\label{ssec:precoding}

Closely related problems to the power control problems discussed above are found in precoding schemes for multiple-input multiple-output (MIMO) multiple-access channels. The instantaneous mutual information of TX $ i $, assuming that multiuser interference can be modeled as a Gaussian random variable, is expressed as
\begin{multline}
 B \log_2 \left\lvert \bm{I}_{ M_\mathrm{ r } } + \rho \bm{H}_i \bm{Q}_i \bm{H}_i^\mathrm{H} + \rho \textstyle\sum_{ j \neq i } \bm{H}_j \bm{Q}_j \bm{H}_j^\mathrm{H} \right\rvert \\
- B \log_2 \left\lvert \bm{I}_{ M_\mathrm{ r } } + \rho \textstyle\sum_{ j \neq i } \bm{H}_j \bm{Q}_j \bm{H}_j^\mathrm{H} \right\rvert,
 \label{eq:mutual_inf}
\end{multline}
where
$ \rho = 1 / N $,
$ \bm{ H }_i \in \mathbb{C}^{ N_\mathrm{ r } \times N_\mathrm{ t } }$ is the channel matrix,
$ \bm{ H }_i^\mathrm{H} $ is the Hermitian transpose of $ \bm{ H }_i $,
$ \bm{ Q }_i $ is a covariance matrix of input signal,
$ M_\mathrm{ t } $ is the number of antennas at every TX,
and $ M_\mathrm{ r } $ is the number of antennas at a single RX.
Belmega et al.\@ \cite{Belmega2010} discussed a game where an input covariance matrix $ \bm{ Q }_i $ is adjusted. Concurrently, Zhong et al.\@ \cite{Zhong2010game} discussed a game where a precoding matrix is adjusted. Since $ \bm{ Q }_i $ is calculated from a precoding matrix, these games are equivalent.

Since (\ref{eq:mutual_inf}) is a coordination-dummy function, this game is an EPG with the system's achievable sum-rate as potential. This game was further discussed in \cite[Section 8]{Lasaulce2011}. Energy efficiency \cite{Zhong2013}, primary-secondary scenario \cite{Zhong2011}, and relay selection \cite{Zhong2014SP} were also discussed. Joint precoding and AP selection in multi-carrier system was discussed in \cite{Mai2015WCOML}.

\section{Transmission Power Assignment Maintaining Connectivity (Topology Control)}
\label{sec:topology}

The primary goal of topology control \cite{Santi2005} is to adjust transmission power to maintain network connectivity while reducing energy consumption to extend network lifetime and/or reducing interference to enhance throughput.

Komali et al.\@ \cite{Komali2008effect} formulated the topology control problem in the TX network model shown in Fig.~\ref{fig:system_models}\subref{fig:model3} as $ \mathcal{G}\game[\label{eq:Komali}] \coloneqq ( \mathcal{ I }, ( \mathcal{ P }_i), ( u\gameref{eq:Komali}_i ) ) $ with $ \mathcal{ P }_i = [0, P_{i,\mathrm{max}}]$ and
\begin{align}
 u{\gameref{eq:Komali}}_i (\bm{p}) &\coloneqq \alpha f_i(\bm{p}) - p_i,
\end{align}
where $\alpha \geq \max_{i} \{ P_{i,\mathrm{max}} \}$, and $f_i(\bm{p})$ is the number of TXs with whom TX $ i $ establishes (possibly over multiple hops) a communication path using bidirectional links. Note that $ f_i ( p_i', \bm{ p }_{ -i } ) \geq f_i ( p_i, \bm{ p }_{ -i } ) $ when $ p_i' > p_i $. This game has been shown to be an OPG with
\begin{align}
 \phi{\gameref{eq:Komali}} (\bm{p}) &= \alpha \sum_{i} f_i (\bm{p}) - \sum_{i} p_i.
\end{align}
Note that the mathematical representation of $ f_i (\bm{ p } ) $ using connectivity matrix \cite{Zavlanos2005} was first proposed in \cite{Moshtagh2010}.

Komali et al.\@ \cite{Komali2009analyzing} also discussed interference reduction through channel assignment, which is seen as a combination of $ \mathcal{G} \gameref{eq:neel} $ and a channel assignment version of $ \mathcal{G}\gameref{eq:Komali}$. They further discussed the impact of the amount of knowledge regarding the network on the spectral efficiency \cite{Komali2010}.

Chu and Sethu \cite{Chu2012} considered battery-operated stations and formulated transmission power control to prolong network lifetime while maintaining connectivity as an OPG. Similar approaches can be found in \cite{Hao2013virtual}, and the joint assignment of transmission power and channels was discussed in \cite{Hao2013distributed}.

Liu et al.\@ \cite{Liu2010QoS,Liu2012} formulated measures for transmission power and sensing range adjustment to enhance energy efficiency while maintaining sensor coverage as an OPG.

\section{Flow and Congestion Control in the Fluid Network Model}
\label{sec:flow}

Ba\c{s}ar et al.\@ \cite{Basar2002,Alpcan2002a} formulated a flow and congestion control game, where each user $ i $ adjusts the amount of traffic flow $ r_i $ to enhance
\begin{align}
 u\gameref{eq:basar}_i ( \bm{ r } ) &\coloneqq \alpha_i \log( 1 + r_i ) - \beta r_i - \frac{ 1 }{ \textit{ capacity }- \sum_i r_i },
 \label{eq:basar_utility}
\end{align}
where $ 1/ ( \mathit{ capacity } - \sum_i r_i ) $ represents the commodity-link cost of congestion. Because (\ref{eq:basar_utility}) is a combination of self-motivated and coordination functions, a game $\mathcal{G}\game[\label{eq:basar}]$ with payoff function $ u\gameref{eq:basar}_i( \bm{ r } ) $ is an EPG \cite{Altman2004,Tsamis2009} with
\begin{align}
 \phi{\gameref{eq:basar}} ( \bm{ r } ) &= \sum_{i} \left( \alpha_i \log ( 1 + r_i ) - \beta r_i \right) - \frac{ 1 }{ \textit{ capacity }- \sum_i r_i }.
\end{align}
The learning process of this game was further discussed by Scutari et al.\@ \cite{Scutari2010}. Other payoff functions for flow control were discussed in \cite{Lin2011,Elias2011}.

\section{Arrival Rate Control for an M/M/1 Queue}
\label{sec:mm1}

Douligeris and Mazumdar \cite{Douligeris1992}, and Zhang and Douligeris \cite{Zhang1992} introduced an M/M/1 queuing game $ \mathcal{G}\game[\label{eq:mm1}] \coloneqq ( \mathcal{ I }, ( \varLambda_i ), ( u\ref{eq:mm1}_i ) ) $, where each user $i$ transmits packets to a single server at departure rate $ \mu $ and adjusts the arrival rate $\lambda_i$ to maximize the ``power'' \cite{Mankin1991}, which is defined as the throughput $\lambda_i$ divided by the delay $\mu - \sum_i \lambda_i$, i.e.,
\begin{align}
 u\gameref{eq:mm1}_i ( \bm{ \lambda } ) &\coloneqq \lambda_i^{ \alpha_i } \left( \mu - \sum_i \lambda_i \right),
\end{align}
where $ \alpha_i\ ( > 0)$ is a factor that controls the trade-off between throughput and delay. Note that this game is a Cournot game (see (\ref{eq:cournot})) when $ \alpha_i = 1 $ for every $ i $.

Gai et al.\@ \cite{Gai2011-14} proved that $ \mathcal{ G }\gameref{eq:mm1} $ is an OPG. Here, we provide a different proof. Because a game with payoff function $ u\game[\label{eq:mm1_dash}]_i( \bm{ \lambda } ) = \alpha_i \log ( \lambda_i ) + \log \left( \mu - \sum_i \lambda_i \right) $
is an EPG, by Theorem \ref{eq:ordinal_transform},  $ ( \mathcal{ I }, ( \varLambda_i ), ( \exp( u\gameref{eq:mm1_dash}_i ) ) ) = ( \mathcal{ I }, ( \varLambda_i ), ( u\gameref{eq:mm1}_i ) ) = \mathcal{ G }\gameref{eq:mm1} $ is an OPG.

\section{Location Update for Mobile Nodes}
\label{sec:location}

\subsection{Connectivity}

Marden et al.\@ \cite{Marden2009} pointed out that the sensor deployment problem (see \cite{Cassandras2005} and references therein), where each mobile node $i$ updates its location $ r_i \in \mathbb{R}^2 $ to forward data from immobile sources to immobile destinations, can be formulated as an EPG. Since the required transmission power to an adjacent node $ j \in \mathcal{ I }_i $ is proportional to the square of the propagation distance, $ \lVert r_i - r_i \rVert $, in a free-space propagation environment, minimizing the total required transmission power problem is formulated as a maximization problem with global objective
\begin{align}
 \phi{\gameref{eq:marden_location}} ( \bm{ r } ) &\coloneqq - \sum_{ i } \sum_{ j \in \mathcal{ I }_i } \frac{ \lVert r_i - r_j \rVert^2 }{ 2 }.
\end{align}
If
\begin{align}
 u\gameref{eq:marden_location}_i ( \bm{ r } ) &= - \sum_{ j \in \mathcal{ I }_i } \lVert r_i - r_j \rVert^2
\end{align}
is used as the payoff function of node $ i $, $\mathcal{G}\game[\label{eq:marden_location}] \coloneqq ( \mathcal{ I }, ( \mathbb{R}^2 ), ( u\gameref{eq:marden_location}_i ) ) $ is equivalent to the consensus game $ \mathcal{G}\gameref{eq:marden_consensus}$.

\subsection{Coverage}
\label{ssec:mobile_sensor}

A sensor coverage problem is formulated as a maximization problem with global objective in continuous form $\phi\game[\label{eq:coverage_continuous}] ( \bm{ s } ) $ \cite{Cassandras2005}
\begin{align}
 \phi{\gameref{eq:coverage_continuous}} ( \bm{ s } ) &\coloneqq \int_{ \Omega } R( r ) \left[ 1 - \prod_{ i } ( 1 - \rho_i ( r, s_i ) ) \right] \, \dd r,
\end{align}
or in discrete form \cite{Murphey2000}
\begin{align}
 \phi{\gameref{eq:coverage_discrete}} ( \bm{ s } ) &\coloneqq \sum_{ r } R( r ) \left[ 1 - \prod_{ i } ( 1 - \rho_i ( r, s_i) ) \right],
\end{align}
where $ \Omega \subset \mathbb{ R }^2 $ is the specific region to be monitored, $ R \colon \Omega \to \mathbb{ R }_{ + } $ is an event density function or value function that indicates the probability density of an event occurring at point $ r \in \Omega $, $ \rho_i \colon \Omega \times \Omega \to [0, 1] $ is the probability of sensor $ i $ to detect an event occurring at $ r \in \Omega $, and $ s_i \in \Omega $ is the location of sensor $ i $. For a summary of coverage problems, we refer the interested reader to \cite{Cardei2005,Cardei2006}.

Arslan et al.\@ \cite{Arslan2007autonomous}
discussed a game where each mobile sensor $ i $ updates its location $ s_i \in \Omega $, treated $ \phi{\gameref{eq:coverage_discrete}} ( \bm{ s } ) $ as potential, and proposed assigning a WLU to each sensor, i.e.,
\begin{align}
 u\gameref{eq:coverage_discrete}_i ( \bm{ s } ) &= \sum_{ r } R( r ) \, \rho_i ( r, s_i ) \prod_{ j \neq i } ( 1 - \rho_j ( r, s_j) ),
 \label{eq:coverage_discrete_utility}
\end{align}
where $ \rho_i ( r, s_i ) \prod_{ j \neq i } ( 1 - \rho_j ( r, s_j) ) $ corresponds to the probability that sensor $ i $ detects an event occurring at $ r $ alone.
Further discussion can be found in \cite{Marden2009} .
We would like to note that $\mathcal{G}\game[\label{eq:coverage_discrete}] \coloneqq ( \mathcal{ I }, ( \Omega ), ( u\gameref{eq:coverage_discrete}_i ) )$ has a similar expression with $ \mathcal{G}\gameref{eq:SALOHA}$. In the same manner in $ \mathcal{G}\gameref{eq:coverage_discrete}$, D\"urr et al.\@ \cite{Durr2011} treated $ \phi{\gameref{eq:coverage_continuous}} ( \bm{ s } ) $ as potential and proposed assigning a WLU
\begin{align}
 u{\gameref{eq:coverage_continuous}}_i ( \bm{ s } ) = \int_\Omega R(r) \, \rho_i(r, s_i) \left[
 \prod_{j \neq i} (1 - \rho_j(r, s_j))
 \right]\, \dd r.
 \label{eq:coverage_continuous_utility}
\end{align}

Zhu and Mart\'{i}nez \cite{Zhu2013} considered mobile sensors with a directional sensing area. Each mobile sensor updates its location and direction. The reward from a target is fairly allocated to sensors covering the target.

Arsie et al.\@ \cite{Arsie2009} considered a game where each node $ i $ attempts to maximize the expected value of the reward. Here, each node $ i $ receives the reward if node $ i $ is the first to reach point $ r $, and the value of the reward is the time until the second node arrives, i.e.,
\begin{align}
 u\gameref{eq:Arsie}_i ( \bm{ s } ) &\coloneqq \int_{ \Omega } R(r) \max \left\{ 0, \min_{ j \neq i } \lVert r - s_j \rVert - \lVert r - s_i \rVert \right\} \, \dd r.
\end{align}
$\mathcal{G} \game[\label{eq:Arsie}] \coloneqq ( \mathcal{ I }, ( \Omega ), ( u\gameref{eq:Arsie}_i ) ) $ was proved to be an EPG.

\section{Channel Assignment to Enhance Coverage for Immobile Sensors}
\label{sec:immobile_sensor}

Ai et al.\@ \cite{Ai2008} formulated a time slot assignment problem for immobile sensors, which is equivalent to a channel allocation problem, as $ \mathcal{ G } \game[\label{eq:Ai}] \coloneqq ( \mathcal{ I }, ( \mathcal{ C }_i ), ( u\gameref{eq:Ai}_i ) )$, where each sensor $ i \in \mathcal{ I } $ selects a slot $ c_i \in \mathcal{ C }_i \coloneqq \{ 1, \dots, K \}$ to maximize the area covered only by sensor $ i $, i.e.,
\begin{align}
 u\gameref{eq:Ai}_i ( \bm{ c } ) &\coloneqq \left\lvert \mathcal{ S }_i \setminus \textstyle\bigcup_{\substack{ j \neq i \\ c_j = c_i } } \mathcal{ S }_j \right\rvert,
\end{align}
where $ \mathcal{ S }_i $ is the sensing area covered by sensor $i$. Game $ \mathcal{G}\gameref{eq:Ai}$ was proved to be an EPG with potential
\begin{align}
 \phi{\gameref{eq:Ai}} ( \bm{ c } ) &= \sum_{ k = 1 }^K \left\lvert \textstyle\bigcup_{\substack{ i \in \mathcal{ I } \\ c_i = k } } \mathcal{ S }_i \right\rvert,
\end{align}
where $ \phi( \bm{ c } ) / K $ corresponds to the average coverage.

To show the close relationship between the payoff functions (\ref{eq:SALOHA_utility}) in the slotted ALOHA game $ \mathcal{G}\gameref{eq:SALOHA}$ and (\ref{eq:coverage_continuous_utility}) in the coverage game $ \mathcal{G}\gameref{eq:coverage_continuous}$, we provide different expressions. Using $ \rho_i ( r ) \coloneqq \indicator{ r \in \mathcal{ S }_i } $, we get
\begin{align}
 u{\gameref{eq:Ai}}_i ( \bm{ c } ) &= \int \rho_i ( r ) \prod_{ j \neq i } ( 1 - \rho_j ( r ) \indicator{ c_j = c_i } ) \, \dd r,
\end{align}
where the surface integral is taken over the whole area. Wang et al.\@ \cite{Wang2012a} further discussed this problem.

Song et al.\@ \cite{Song2011distributed} applied the coverage game to camera networks. Ding et al.\@ \cite{Ding2012} discussed a pan-tilt-zoom (PTZ) camera network to track multiple targets. Another potential game-theoretic PTZ camera control scheme was proposed in \cite{Hatanaka2013CDC}, motivated by natural environmental monitoring. Directional sensors were discussed in \cite{Li2014}. The form of payoff functions is similar to (\ref{eq:coverage_discrete_utility}).

Until now, each immobile sensor was assumed to receive a payoff when it covered a target alone. Yen et al.\@ \cite{Yen2013} discussed a game where each sensor receives a payoff when the number of sensors covering a target is smaller than or equal to the allowable number. Since this game falls within a class of CGs, it is also an EPG.

\section{Conclusions}
\label{sec:conclusion}

We have provided a comprehensive survey of potential game approaches to wireless networks, including channel assignment problems and transmission power assignment problems. Although there are a variety of payoff functions that have been proven to have potential, there are some representative forms, e.g., BSI games and congestion games, and we have shown the relations between representative forms and individual payoff functions. We hope the relations shown in this paper will provide insights useful in designing wireless technologies.

Other problems that have been formulated in terms of potential games are found in routing \cite{Altman2007,Zeydan2012,Liang2014,Xiao2014,Secci2011,Secci2013}, BS/AP selection \cite{Singh2009,Lin2011ICC,Malanchini2013,Lin2014,Tseng2013}, cooperative transmissions \cite{Ng2010,Al-Tous2015WCOM}, secrecy rate maximization \cite{Alvarado2014}, code design for radar \cite{Piezzo2013}, broadcasting \cite{Chen2013gamebased}, spectrum market \cite{Korcak2012}, network coding \cite{Reddy2010INFOCOM,Marden2012price,Ramaswamy2014}, data cashing \cite{Li2012game}, social networks \cite{Chen2014social}, computation offloading \cite{Chen2015PDS}, localization \cite{Jia2013}, and demand-side management in smart grids \cite{Ibars2010SMARGRIDCOMM,Wu2011GLOBECOM}.

\section*{Acknowledgments}
This work was supported in part by JSPS KAKENHI Grant Numbers 24360149, 15K06062. The author would like to acknowledge Dr.\ Takeshi Hatanaka at Tokyo Institute of Technology for his insightful comments on cooperative control and PTZ camera control. The author acknowledges Dr.\ I. Wayan Mustika at Universitas Gadjah Mada, and Dr.\ Masahiro Morikura and Dr.\ Takayuki Nishio at Kyoto University for their comments.

\begin{IEEEbiographynophoto}{Koji Yamamoto}
received the B.E.\ degree in electrical and
electronic engineering from Kyoto University in 2002, and the M.E.\ and
Ph.D.\ degrees in informatics from Kyoto University in 2004 and 2005,
respectively.  From 2004 to 2005, he was a research fellow of the Japan
Society for the Promotion of Science (JSPS).  Since 2005, he has been
with the Graduate School of Informatics, Kyoto University, where he is
currently an associate professor.  From 2008 to 2009, he was a visiting
researcher at Wireless@KTH, Royal Institute of Technology (KTH) in
Sweden.  His research interests include the application of game theory,
spectrum sharing, and in-band full-duplex communications. He received the PIMRC 2004 Best
Student Paper Award in 2004, the Ericsson Young Scientist Award in 2006,
and the Young Researcher's Award from the IEICE of Japan in 2008.  He is
a member of the IEEE.  
\end{IEEEbiographynophoto}

\end{document}